\documentclass[fleqn,usenatbib]{mnras} 

\usepackage{newtxtext,newtxmath}

\usepackage[T1]{fontenc}
\usepackage{ae,aecompl}
\usepackage{float}
\usepackage{bold-extra}
\restylefloat{table}

\usepackage{graphicx}	
\usepackage{amsmath}	
\usepackage{amssymb}	

\usepackage{array,multirow}
\usepackage{xcolor,colortbl}
\usepackage{fancybox}
\usepackage{framed}
\usepackage{transparent}
\usepackage{wrapfig}
\usepackage{soul}
\usepackage{amsmath}
\usepackage{textcomp}
\usepackage{listings}
\usepackage[toc,page]{appendix}
\usepackage[draft]{minted}
\usepackage{hyperref}
\usepackage{caption}
\usepackage{subcaption}
\usepackage{verbatim}
\usepackage[version=4]{mhchem}
\usepackage{nicefrac}
\long\def\/*#1*/{}
\DeclareMathOperator\erf{erf}

\newcommand{\rev}[1]{\textcolor{black}{#1}}

\def\fig{Fig.\,}

\def\sec{Section~}
\def\tab{Table\,}

\def\msunh{\,h^{-1}\rm{\,M_{\odot}}}

\def\kms{\rm{\,km\,s^{-1}}}

\newcommand{\be}{\begin{equation}}
\newcommand{\ee}{\end{equation}}
\def\aap{A\&A}
\def\apj{ApJ}

\def\apjl{ApJ}
\def\mnras{MNRAS}
\def\araa{ARA\&A}
\def\aj{AJ}

\def\apjs{ApJS}

\def\pasp{Publ. Astr. Soc. Pac.}

\def\pasj{Pub. Astron. Soc. Japan}

\title[Dust Evolution in Galaxy Cluster Simulations]{Dust Evolution in Galaxy Cluster Simulations }

\author[E. Gjergo et al.]{
Eda Gjergo$^{1,2}$\thanks{E-mail: gjergo@oats.inaf.it (EG)}, Gian Luigi Granato$^{2,3}$, 
Giuseppe Murante$^{2,3}$, \\~\\
{\rm {\LARGE
Cinthia Ragone-Figueroa$^{3,2}$,
Luca Tornatore$^{2}$,
Stefano Borgani$^{1,2}$}}
\\ \\
$^{1}$ Dipartimento di Fisica dell' Universit\`a di Trieste, Sezione di Astronomia, via Tiepolo 11, I-34131 Trieste, Italy\\
$^{2}$ INAF, Osservatorio Astronomico di Trieste, via Tiepolo 11, I-34131, Trieste, Italy\\
$^{3}$ Instituto de Astronom\'ia Te\'orica y Experimental (IATE), Consejo Nacional de Investigaciones Cient\'ificas
y T\'ecnicas de la Rep\'ublica \\
Argentina (CONICET), Observatorio Astron\'omico, Universidad Nacional de C\'ordoba, Laprida 854, X5000BGR C\'ordoba, Argentina\\
}

\date{Accepted XXX. Received YYY; in original form ZZZ}

\pubyear{2018}

\begin{document}
\label{firstpage}
\pagerange{\pageref{firstpage}--\pageref{lastpage}}
\maketitle
\defcitealias{43planck16}{Planck-XLIII}
\defcitealias{gutierrez17}{G-LC17}
\defcitealias{mcgee10}{MG-B10}
\begin{abstract}
We implement a state-of-the-art treatment of the processes affecting the production and Interstellar Medium (ISM) evolution of carbonaceous and silicate dust grains within SPH simulations. We trace the dust grain size distribution by means of a two-size approximation. We test our method  on zoom-in simulations of four massive ($M_{200} \geq 3 \times 10^{14} M_{\odot}$) galaxy clusters. 
We predict that during the early stages of assembly of the cluster at  $z \gtrsim 3$,  where the star formation activity is at its maximum in our simulations, the proto-cluster regions are rich in dusty gas. Compared to the case in which only dust production in stellar ejecta is active, if we include processes occurring in the cold ISM,the dust content is enhanced by a factor $2-3$.
However, the dust properties in this stage turn out to be significantly different from those observationally derived for the {\it average} Milky Way dust, and commonly adopted in calculations of dust reprocessing. We show that these differences may have a strong impact on the predicted spectral energy distributions. At low redshift in star forming regions our model reproduces reasonably well the trend of dust abundances over metallicity as observed in local galaxies. However we under-produce by a factor of 2 to 3 the total dust content of clusters estimated observationally at low redshift, $z \lesssim 0.5$ using IRAS, Planck and Herschel satellites data. This discrepancy does not subsist by assuming a lower sputtering efficiency, which erodes dust grains in the hot Intracluster Medium (ICM).

\end{abstract}



\section{Introduction}
A significant fraction of metals present in the interstellar medium (ISM) is {\it depleted} from the gas phase and locked into small solid  particles, the {\it cosmic dust}.  The size of these particles is distributed over a broad range,  from a few tens of \AA\ up to a few $\mu$m. In the Milky Way, about 50\% of the metal mass, or about 1\% of the ISM mass, is in dust. Theoretical works \citep[e.g.][]{dwek98} predict the first percentage to be roughly constant, and as a consequence the second is approximately proportional to the ambient gas metallicity $Z_{gas}$. This estimate is confirmed by observations of metal-rich galaxies, but it is subject to a broad scatter. On the other hand, low metallicity dwarf galaxies, having a low dust-to-gas (DtG) mass ratio, deviate substatially \rev{\citep[e.g.][]{galametz11,remy14}}. 
As for the detailed chemical composition of dust, as a first approximation it is accepted that dust consists of two major chemical classes: one carbon-based, and another named ``astronomical silicates", composed of essentially four elements, O, Si, Mg and Fe. This is supported by depletion and dust-reprocessing studies \citep[e.g.][and references therein]{draine03,jenkins09}. 

Despite this deceivingly reassuring summary of cosmic dust properties, it is clear that the situation is much more complex \citep[e.g.][and references therein]{jones13}. All the above mentioned properties of dust represent only the average at late cosmic times. However, observations show significant deviations, both from galaxy to galaxy in the local Universe, and within different environments of the Milky Way itself.
Moreover, observations suggest substantial differences at early cosmic times. From a theoretical perspective, this is all but surprising. Indeed, dust grains constitute a living component of the ISM. Once dust grain seeds are produced,  mostly in stellar outflows, they are subject to several evolutionary processes in the ambient gas, whose effectiveness depends on the physical and chemical gas conditions, as well as on the properties of the grains themselves. These processes, which we will describe in Section \ref{sec:dusfor}, alter the abundance, chemical composition and size distribution of dust grains.

Most galaxy properties cannot be described accurately without accounting for dust. For instance, the chemical species that dust depletes are key ISM coolants. Moreover, dust surfaces catalyze the formation of molecular species such as $\ce{H_2}$ \citep[e.g.][]{barlow76}. $\ce{H_2}$ is the primary constituent of molecular clouds (MCs), the star formation sites. Among all dust effects, the best studied is the dust reprocessing of radiation emitted by astrophysical objects. Therefore, dust is of paramount importance in interpreting observations.
Dust absorbs efficiently stellar (or AGN) UV and optical light. The absorbed energy is thermally re-emitted in the IR, mostly at $\lambda> \mbox{a few}\ \mu$m, with a peak at about 100 $\mu$m. In the local Universe, dust contributes only to less than 1\% of the ISM mass. Despite being so scarce, it reprocesses about 30\% of all the photons emitted by stars and AGNs \citep{soifer1991}. The reprocessed fraction increases as a function of the specific star formation activity of galaxies \citep{sanders1996}. This is because in star forming objects, the primary radiative power originates from young stars, which are close or still embedded in their parent MCs \cite[][]{silva98,granato00,charlot00}. Thus, IR turns out to be a very good observational tracer for star formation \citep[eg][]{kennicutt12}. It is worth noticing that the absorption and scattering cross sections of grains depend not only on their composition but also, and strongly, on their radius $a$\footnote{Usually dust grains are simply approximated by spheres.}. Therefore, reliable galactic SED models must take dust size distribution into account \citep{silva98}.

In this work, we implement within the {\footnotesize GADGET-3} SPH code a state-of-the-art treatment of the processes affecting the production, evolution and destruction of carbonaceous and silicate dust grains. We model the size distribution of dust grains by means of the two-size approximation developed by \cite{hirashita15}. His work  demonstrates that it is possible to reproduce the broad results of a full grain-size treatment just considering only two well-chosen representative sizes. The computation of this solution is not very demanding and therefore it is well-suited for cosmological simulations. Moreover, the method can be generalized in the future to increase the number of grain sizes.  

While the {\footnotesize GADGET-3} code is suitable for simulations of galaxy formation, we apply it here to zoom-in simulations of two massive ($\sim 10^{15} M_{\odot}$) and two smaller ($\sim 5 \times 10^{14} M_{\odot}$) galaxy clusters. We are mainly interested in the high redshift stages,  where the star formation activity is at its maximum, and the proto-cluster regions are rich of cold dusty gas. The first aim of our dust evolution model is to couple it in the near future with post-processing radiative transfer computations. We plan to replace the uncertain assumptions on the dust content, chemical composition, and size distribution, with estimates derived from the computation of these properties in the simulated ISM. The presence of dust production and evolution will allow, at a second stage, to account for the role of dust as a catalyst for the formation of molecules, as well as the  impact of gas heating and cooling due to dust \citep{montier04}.

A few groups have successfully included some aspects of the evolution of dust content of the ISM within hydrodynamical simulations \rev{\citep[e.g.][]{bekki13,mckinnon16,mckinnon17,aoyama17,hou17}}.\rev{Another possibility is to investigate the problem by means of post-processing computations 
\citep[e.g.][]{zhukovska16} or within semi-analytic models \cite{popping16}.}
However the present work provides a more comprehensive description of dust as we predict both the size distribution and the chemical composition of dust grains {\it at the same time}, self consistently with a full treatment of the chemical evolution of the ISM. Moreover, while the focus of the aforementioned papers was on galactic disks, our work represents the first attempt to trace the evolution of dust component in simulations of galaxy clusters.

We follow \cite{dwek98} in using a notation for metallicity that specifies when we refer to metals in the gas form $Z_{gas}$, metals in the solid form, i.e. dust $Z_{dust}$, or the sum of the two $Z_{tot}$.
This is not a standard convention as in other works dealing with the evolution of the dust in the ISM, the symbol $Z$ has been employed to refer to the {\it total} metal fraction (metallicity) of the ISM \citep[e.g.][]{calura08,hirashita15}, including both metals in gas and metals locked up in solid state grains.

This paper is organized as follows: in \sec\ref{sec:model} we describe the model. We give a brief overview of our cosmological simulation set and we present our implementation of the dust evolution model. In \sec\ref{sec:results} we describe and analyze the outcome of said dust model for a variety of global and internal properties. We also offer a preliminary data comparison. \sec\ref{sec:conclusions} reports the conclusions of the present work.

\section{Model} \label{sec:model}
After implementing a treatment of dust evolution, in this work we re-run  and analyze four zoom-in simulations for the formation of massive galaxy clusters, taken from a sample already presented in a number of papers by our group. We recall briefly in this section  the main aspects of these simulations, while we refer the reader in particular to \cite{ragone13} and \cite{ragone18}
for further numerical and technical details. The modifications introduced to take into account the evolution of dust in the ISM are fully described in Section \ref{sec:dusfor}. 

\subsection{The parent sample of simulated clusters}
\label{sec:numerical}
The full set of initial conditions describe 29 zoomed-in Lagrangian regions, selected from a  parent DM-only and low resolution simulation of a 1 $h^{-1}$ Gpc box. The regions are built to safely contain the 24 dark matter halos with $M_{200}>8\times 10^{14}\, \msunh$ and 5 smaller ones with $M_{200}=[1-4]\times 10^{14}\, \msunh$\footnote{$M_{200}$ ($M_{500}$) is the mass enclosed by a sphere whose mean density is 200 (500) times the critical density at the considered redshift.}. We adopt the following cosmological parameters: $\Omega_{\rm{m}} = 0.24$, $\Omega_{\rm{b}} = 0.04$, $n_{\rm{s}}=0.96$, $\sigma_8 =0.8$ and $H_0=72\,\kms$\,Mpc$^{-1}$.

These regions are re-simulated with a custom version of the {\footnotesize GADGET-3} code~\cite[][]{springel05} to achieve better resolution and to include the baryonic physics. The mass resolution for the DM and gas particles is $m_{\rm{DM}} = 8.47\times10^8 \, \msunh$ and $m_{\rm{gas}} =1.53\times10^8\, \msunh$, 
respectively. \rev{Note that the latter value is the {\it initial} mass of gas particles. However, as described in Section \ref{sec:unresolved}, during the simulation these particles can decrease their mass by spawning up to 4 stellar particles, to simulate star formation. On the other hand, they can also gain mass due to gas ejecta produced by neighborhood stellar particles. Thus the mass of each gas particle evolve with time. At $z=0$, the 5\% and 95\% percentiles of their mass distribution turns out to be 0.6 and 1.06 the initial value respectively.}

For the gravitational force, a
Plummer-equivalent softening length of $\epsilon = 5.6\, h^{-1}$\,kpc is used for DM and gas particles, whereas $\epsilon = 3\, h^{-1}$\,kpc for black hole and star particles. The DM softening length is kept fixed in comoving coordinates for $z>2$ and in physical coordinates at lower redshift. To treat hydrodynamical forces, we adopt the SPH formulation by \cite{beck16}, that includes artificial thermal diffusion and a higher-order interpolation kernel, which improves the standard SPH performance in its capability of treating discontinuities and following the development of gas-dynamical instabilities.

\subsubsection{Unresolved physics}
\label{sec:unresolved}
Our simulations include a treatment of all the unresolved baryonic processes
usually taken into account in galaxy formation simulations. For details on the adopted implementation of cooling, star formation (SF), and associated feedback, we refer the reader to \cite{planelles14}. In brief, the model of SF is an updated version of the implementation by~\cite{springel03}, in which gas particles with a density above $0.1\,$cm$^{-3}$ and a temperature below $T_{MPh}=5\times 10^5$\,K are classified as multiphase. Multiphase particles consist of a cold and a hot-phase, in pressure equilibrium. The cold phase is the star formation reservoir. At each timestep, multiphase particles can stochastically spawn non-collisional star particles, each representing a simple stellar population, with expectation value consistent with the star formation rate. The latter is computed on the basis of the physical state of the particle. Each star-forming gas particle can generate up to $N_g$ (=4 in our simulations) star particles. The $N_g$-th event, when it occurs, consists in the complete conversion into a star particle.
\rev{We implement kinetic feedback following the prescription by \cite{springel03}, in which a multiphase star-forming particle has a probability to be loaded into galactic outflows (we assume $v_w = 350 \mbox{km s}^{-1}$ for the outflow velocity). This probability is set to generate on average an outflow rate proportional to the star formation rate, with a proportionality factor of 2}.

\rev{
A full account of the AGN feedback model can be found in Appendix A of \cite{ragone13}, with some modifications described in \citep{ragone18}. We refer the reader to those papers for the numerical details.  In short the BHs are represented by collision-less particles, seeded at the center of a DM halos when they become more massive than $M_{th}=2.5 \times 10^{11} h^{-1}$ M$_\odot$ and do not already contain a SMBH. The initial BH mass at seeding is $M_{seed}=5 \cdot 10^6 h^{-1}$ M$_\odot$. 
We reposition at each time-step the SMBHs particle at the position of the nearby particle, of whatever type, having the minimum value of the gravitational potential within the gravitational softening.
The SMBH grows with an accretion rate given by the minimum between the $\alpha$-modified  Bondi accretion rate and the Eddington limit. The formula for the $\alpha$-modified Bondi accretion rate:
\begin{equation}
  \dot{M}_{Bondi,\alpha} = \alpha \,   \frac{4 \pi G^2M^2_{BH} \rho}{\left(c^2_s+v^2_{BH}\right)^{2/3}}
\label{eq:bondi}
\end{equation}
is applied separately to the hot and cold gas components.
The threshold between the two components is set at $T=5\times 10^5$ K, and the adopted values of the fudge factor $\alpha$ are 100 and 10 respectively. This distinction between cold and hot accretion modes has been inspired by the result of high resolution AMR simulations of the gas flowing to SMBHs \citep{gaspari13}.
}
\rev{BHs particles merge when they fall within the gravitational softening.
The accretion of gas onto the SMBH produces an energy determined by the radiative efficiency parameter $\epsilon_r$. Another parameter $\epsilon_f$ defines the fraction of this energy that is thermally coupled to the surrounding gas. We calibrated these parameters to reproduce the observed scaling relations of SMBH mass in spheroids. Here we set $\epsilon_r=0.07$ and $\epsilon_f=0.1$,  Finally, we assume a transition from a {\it quasar mode} to a {\it radio mode} AGN feedback when the accretion rate becomes smaller than a given limit, $\dot{M}_{BH}/\dot{M}_{Edd} = 10^{-2}$. In this case, we increase the feedback efficiency $\epsilon_f$ to 0.7.
}

\subsection{The model for chemical enrichment}
\label{Zprod}
Stellar evolution and metal enrichment follow 
\cite{tornatore07}. The production of heavy elements considers separately contributions from Asymptotic Giant Branch (AGB) stellar winds, Type Ia Supernovae (SNIa) and Type II Supernovae (SNII), which is required also when implementing the production of dust from the different stellar channels (Section \ref{dsynth}). Whereas all three types of stars contribute to the chemical enrichment, only SNIa and SNII provide thermal feedback.  In addition, as described in~\cite{springel03}, kinetic feedback from SNII is implemented as galactic outflows with a wind velocity of $350 \kms$.  The initial mass distribution for the star population is described by  the initial mass function of~\cite{chabrier03}. We assume the mass-dependent lifetimes of~\cite{padovani93} to account for the different time-scales of stars of different masses.

To estimate the production of metals due to the evolution of stellar particles, we consider different sets of stellar yields, as detailed in \cite{biffi17}. We trace the production and evolution of 15 chemical elements: H, He, C, Ca, O, N, Ne, Mg, S, Si, Fe, Na, Al, Ar and Ni. Consistently, metallicity-dependent radiative cooling of gas is calculated by taking into account the contribution of these 15 chemical species.  

The code do not include any special treatment of metal diffusion, although it accounts for the spreading of metals (both in gaseous and dust form, see below) from star particles to the surrounding gas particles by using the same kernel of the SPH interpolation.  This is mainly done in order to avoid a noisy estimation of metal-dependent cooling rates. Therefore heavy elements and dust can be spatially distributed after that only via dynamical processes involving the metal-rich gas. 

\subsection{The test subsample for this work} \label{sec:simclus}
To test our implementation of dust evolution we select 4 of the 29 zoomed-in regions. Two of them, dubbed D2 and D3, belong to the small mass sample of 5 regions discussed above, while the other 2, D1 and D6, are in the large mass sample.
Their respective $M_{200}$ at $z = 0$ are  $5.4 \mbox{, } 6.8 \mbox{, } 11.8\mbox{, } \mbox{and } 17.5 \times 10^{14} M_{\odot}$. Most of the following analysis refers to the main cluster of region D2. 


\subsection{Dust formation and evolution}
\label{sec:dusfor}

We implemented within our version of {\footnotesize GADGET-3} a treatment of dust evolution  similar to that introduced by \cite{hirashita15} and \cite{aoyama17}, with some modifications.  We anticipate them here for the sake of clarity, before detailing our approach in the rest of this section: (i) We take advantage of a detailed treatment of chemical evolution (Section \ref{Zprod}), which tracks individually several heavy elements instead of just the global metallicity. Therefore we can follow separately the fate of the two distinct dust chemical phases which are believed to populate the ISM, namely carbonaceous and silicate grains; (ii) rather than considering just the instantaneous ISM dust pollution due to SNII explosions,  we are able to properly compute also the delayed  effect of SNIa explosions and of  AGB stars winds; (iii) we take into account also the dust erosion caused by thermal sputtering, which becomes dominant in the hot intra-cluster medium (ICM). 

As for the chemical composition of dust, we adopt the standard view, supported by depletion and radiative studies \citep[e.g.][and references therein]{draine03,jenkins09}, that dust composition is dominated by C, O, Si, Mg and Fe. These elements are believed to be organized in two major chemical classes, one composed mostly by carbon  and another one composed by "astronomical silicates", comprising the four latter elements. More specifically, we accept that the former one is reasonably well  approximated for our purposes by pure C grains, and the latter one by Olivine MgFeSiO$_4$ grains \citep{draine03}. We explicitly note that the majority of radiative transfer computations in astrophysical dusty media are done adopting optical properties of dust grains calculated according to these assumptions. However it is also worth pointing out that our treatment can be immediately adapted to different dust composition models, as long as the relevant elements are included in the chemical evolution treatment described in Section \ref{Zprod}.

As in \cite{aoyama17}, we account for the grain size using the two-size approximation proposed by \cite{hirashita15}, which has been derived from the full treatment of grain size distribution put forward by \cite{asano13}. In the two size approach, just  two discrete size  populations are considered to represent the whole continuum range of  grain radii $a$. We refer to the two populations as large and small grains. This is justified by the fact that the various processes affecting dust population in the ISM act differently in these two representative populations.
The boundary between them is set by \cite{hirashita15} at $a\simeq 0.03 \, \mu$m.
We thus track four types of dust grains: large and small carbonaceous grains, and large and small silicates grains.

It is worth pointing out that specifying the grain size distribution, albeit minimal, has a two-fold effect. It allows a more realistic treatment of the evolution of dust population, and it is also useful for post-processing computations involving radiative transfer that we plan to include in the next future. Indeed, the optical properties of grains strongly depends on their size. The two-size approximation allows for such a grain size treatment without affecting too heavily computation times and memory requirements. In the present application the overhead of computing time arising from the inclusion of dust evolution is limited to less than 20 per cent.

\rev{This two size approximation has been already adopted in previous simulations works \citep{hou16,hou17,chen18} which also tracked separately carbon and silicate dust. Note however that their simulations used a simplified treatment of chemical evolution, the Instantaneous Recycling Approximation (IRA). As a consequence, they included only SNII for the primary production of dust grains from stars, neglecting both AGB and SNIa. According to our results, based on a full treatment of chemical evolution, the SNII channel turns out to be insufficient to properly predict the relative abundance of silicate and carbon grains, and thus the radiative effect of dust (see \ref{sec:evol}}). 

Previous studies have identified the main processes that should be taken into account to understand the dust content of galaxies. The life cycle of grains begins with {\it dust production} in the ejecta of stars, including the relatively quiescent winds of AGB stars,  but also  SNII and SNIa explosion. It can be assumed that dust production affects directly only large grains. Indeed, SNae shocks are more efficient at the destruction of small grains than of large grains, therefore over chemical evolution timescales only the latter will survive \citep{nozawa07,bianchi07}). As for AGB winds, their infrared SEDs suggest that the typical grain size is skewed toward large grains \citep{groenewegen97,gauger99}. These findings are supported by expectations of models including dust formation in AGB winds \citep[e.g.][]{winters97}. 

Once injected into the ISM, the grain population is subject to several evolutionary processes. 
Metals from the surrounding gas can deposit on the grain surface, causing the grains to gain mass, in a process named {\it accretion} \citep{dwek98,hirashita99}. An important role of this process has been invoked to explain the remarkable amount of dust in high-$z$ quasars and starbursts \citep[e.g.][and references therein]{michalowski10,rowlands14,nozawa15}, which cannot be accounted by stellar production only. Being  a surface process, it acts more efficiently on small grains, which have larger total surface per unit mass. However, accretion is not efficient enough to increase significantly a small grain's radius. Its effect is just to increase the amount of small grains, while the evolution of small grains into large grains by accretion is safely negligible.
Grain-grain collision can either result in grain {\it coagulation} or {\it shattering}. The former process dominates in dense ISM, and can be regarded as a destruction mechanism for small grains and a formation mechanism for large ones. The latter mechanism wins in the diffuse ISM, leading to the formation of small grains and destruction of large ones. 

Sputtering, also a surface process, consists of grain erosion due to collisions with energetic ions. The eroded atoms are given back to the gas phase. It occurs both when grains are swept by SNae shocks, and when they are subject to the harsh ion collisions in the hot ICM. On top of SNae dust destruction, we include thermal sputtering in the ICM, which is important only when grains are surrounded by hot gas $T\gtrsim 10^6$ K. \cite{hirashita15} and \cite{aoyama17} did not consider ICM thermal sputtering, because they were interested only in galactic ISM. We refer to this process simply as {\it sputtering}, whether we call the former one  {\it SNae destruction}. 

\begin{figure*}
	\centering
	\includegraphics[width= 0.8\textwidth]{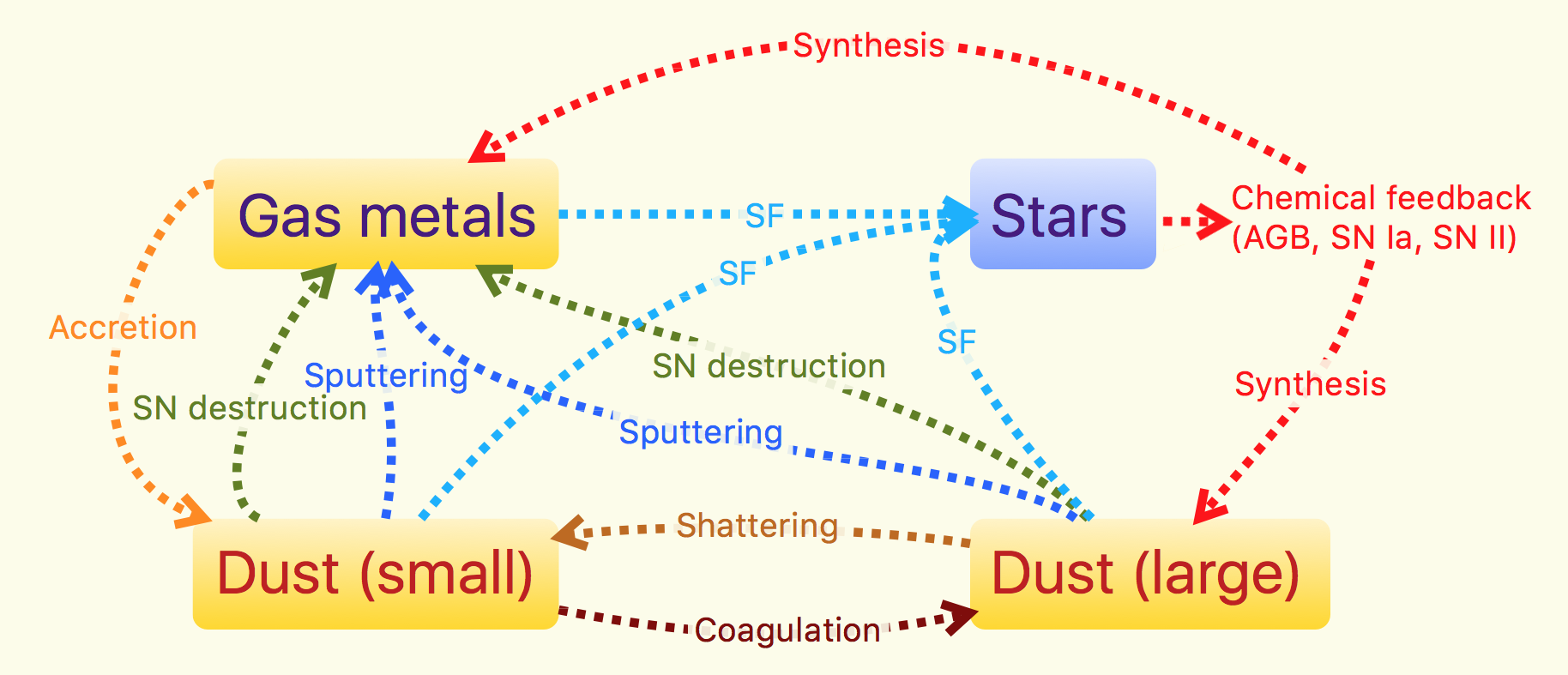}
    \caption{Diagram of mass flows due to dust processes. Boxed in yellow are gas metals, large and small dust, which belong to the gas particle structure. Boxed in blue is the star particle, responsible for the production of metals in the subgrid chemical evolution through AGB winds, SN Ia, and SN II. Star particles spread metals to the surrounding gas particles by enriching gas metals and large dust alike. When a gas particle is ready to form a star particle, all 3 of the gas particle metal channels are depleted in proportion, in favor of the star particle. SN destruction and sputtering subtract metals from both dust channels and enrich the gas metals.}
    \label{fig:dustmap}
\end{figure*}

Needless to say, when gas is turned into stars, its dust content is proportionally subtracted from the ISM. This event is often referred to as {\it astration}.
A visual summary of how the various processes act on the  components of a galaxy is provided in Figure \ref{fig:dustmap}.

We implement this sequence of processes at the sub-grid level. 
A variable fraction of every metal produced by star particles is  given to neighbor gas particles in the form of solid state dust rather than gas. Accordingly, we modified the gas particle structure to trace two extra vectors along with the gas metals: one for large and one for small grains. The three vectors have a length equal to the number of chemical species we trace, which are 15 in this paper. Thus a certain fraction of the mass of each SPH "gas" particle, never exceeding a few \%, is actually used to represent metals which are locked up in solid grains. We still refer to them simply as SPH gas particles.

For a SPH gas particle of total mass $M_{gas}$, the time derivatives of the dust mass content in the two size populations of small grains $M_{d,S}$ and large grains $M_{d,L}$  can be expressed as follows:

\begin{equation}
\begin{split}
    \frac{d {M_{d,L}}}{dt} & =     \frac{d {M_{p*}}}{dt}   - \frac{M_{d,L}}{\tau_{sh}} + \frac{M_{d,S}}{\tau_{co}}  - \frac{M_{d,L}}{\tau_{SN,L}} - \frac{M_{d,L}}{\tau_{sp,L}}- \frac{M_{d,L}}{M_{gas}} \psi \\
    \frac{d { M_{d,S}}}{dt} & = \frac{M_{d,S}}{\tau_{acc}} + \frac{M_{d,L}}{\tau_{sh}} - \frac{M_{d,S}}{\tau_{co}} - \frac{M_{d,S}}{\tau_{SN,S}} - \frac{M_{d,S}}{ \tau_{sp,S}} - \frac{M_{d,S}}{M_{gas}}\psi  
\end{split}
\end{equation}
where $\frac{d {M_{p*}}}{dt}$ is the dust enrichment arising from production by stars, and  local timescales for the various dust processes occurring in the ISM are introduced: $\tau_{sh}$ for shattering, $\tau_{co}$ for coagulation, $\tau_{SN,L}$ and $\tau_{SN,S}$ for SN shock destruction, $\tau_{sp}$ for sputtering and $\tau_{acc}$ for accretion.  The last term of each equation represents the dust mass loss due to star formation (astration), specifically $\psi$ is the star formation rate. \rev{These equations are applied separately for each element entering into the dust grains. This implies that if a  gas particle has gas-phase metals for accretion of silicates but not gas phase C (a quite unlikely possibility, in particular at our relatively coarse resolution), only the mass of small silicates is affected by accretion,  but not that of C grains.}
In the next subsections we give details on how each of these terms is computed in our simulations.

\subsubsection{Dust production by stars} \label{dsynth}
We adapt to each simulation star particle the formalism for dust production by stars introduced by \citet{dwek98}, in the context of {\it monolithic} galaxy formation models. However, we introduce an important modification related to our specific assumption for the chemical composition of silicate grains.

Dust is produced by the same stellar enrichment channels which are responsible of metals production, namely AGB winds, SNIa, and SNII. We assume that a certain fraction of the elements concurring to dust formation is originally given to the ISM locked to solid dust particles rather than in gaseous form. These elements are C, Si, O, Mg, and Fe. This is because, as motivated at the beginning of Section \ref{sec:dusfor}, we consider carbon and silicate dust, and we approximate silicates with olivine MgFeSiO$_4$. \rev{In this exploratory work we maintain for simplicity the assumption that, for each stellar channel the fraction dust elements that condense to dust does not depend on stellar mass nor on metallicity. This approach, introduced in one-zone models by \cite{dwek98}, has been adopted by many later computations  \citep[e.g.][]{calura08,hirashita15,aoyama17}. However we point out that there are several works that studied how dust
condensation efficiencies actually depend on stellar mass, metallicity and ambient gas density \citep[see e.g.][and references therein]{nozawa07,schneider14}. In the future it will be relatively straightforward to incorporate the results of such computations into our formalism.}

\subsubsection{AGB stars}
As for AGB winds, following \citet{dwek98} we assume that the formation of carbon or silicate dust is mutually exclusive, and depends on the C/O number ratio in the ejecta. When C/O$>1$, all the oxygen is tied up in CO molecules and only carbon dust is formed. If instead C/O$<1$, all the carbon is consumed to form CO molecules, while the excess oxygen will combine with Si, Mg and Fe to produce mainly silicate grains. This view is broadly supported by many observations \citep[e.g.][]{kastner93}, and implies the assumption that AGB ejecta are mixed at microscopic level, so that the maximum possible amount of CO is formed.

Let $M_{ej,C}^{AGB}$ and $M_{ej,O}^{AGB}$ be the C and O masses ejected by a star particle through the AGB enrichment channel during a given time-step. 
Carbon (silicate) grains form when $M_{ej,C}^{AGB}$ is greater (smaller) than $  0.75 M_{ej,O}^{AGB}$, where 0.75 is the ratio between O and C atomic weights. Due to the C mass lost to CO, the mass of carbon dust produced during the time-step is given by
\begin{equation}
M_{dust,C}^{AGB}=
\max\left[\delta_{AGB,C}\left( M_{ej,C}^{AGB} - 0.75 \, M_{ej,O}^{AGB} \right)\, , \, 0 \right] 
\end{equation}
where $\delta_{AGB,C}$ is a {\it condensation efficiency} of carbon grains in AGB winds, which we set equal to 1, as in \cite{dwek98}  \citep[see also][]{calura08}.

If $M_{ej,C}^{AGB} <  0.75 M_{ej,O}^{AGB}$, silicate grains are instead produced. \cite{dwek98} estimated the masses of the elements going into silicate dust under the simple assumption that for each atom of them, a single O atom will go into dust as well (see his equation 23). While we have implemented in our code also his formulation, we prefer a more specific assumption on the chemical composition of silicates produced by stars, which leads us to different formulae. As expected we have verified with our code that \cite{dwek98} approach leads to the production of 'silicate grains' featuring very variable mass ratios between the various elements, and typically very different from those implicitly assumed by most radiative transfer computations. 
This is particularly true for the SNIa channel, since these stellar events produce mostly iron, much in excess to the quantity that can be bound directly with O in the SNae ejecta. When we use the formalism by \cite{dwek98}, dust grains reach a mass fraction for Fe which is double that of olivine. Mg is under-abundant by a factor of 3. Our method on the other hand, described below,  preserves the olivine-like mass fractions in the four metal components at all redshifts.

Generalizing the notation introduced above, we indicate as $M_{ej,X}^{AGB}$ the masses ejected by a star particle, through the AGB enrichment channel and during a given time-step, in form of the X element which enters into the olivine compound. X stands then for Mg, Fe, Si and O. We indicate 
$N_{mol,sil}^{AGB}$ as the number of "molecules" of MgFeSiO$_4$ that can be formed over the time-step. This is set by the less abundant element, taking into account how many atoms of each element $N_{ato}^X$ enter into the compound (1 for Mg, Fe, Si, and 4 for O). Then
\begin{equation}
\label{eq:nmol}
N_{mol,sil}^{AGB} = \delta_{AGB,sil} \min\limits_{X  \in \{Mg,Fe,Si,O\}} \left(\frac{M_{ej,X}^{AGB}}{\mu_X \, N_{ato}^X}\right)
\end{equation}
where $\mu_X$ is  the atomic weight of the X element,
and we have introduced an efficiency factor of condensation for silicate grains $\delta_{AGB,sil}$, set to 1 in our reference model. 
Therefore, the mass of the X element locked into silicate grains is given by
\begin{equation}
\label{eq:mdust}
M_{dust,X}^{AGB}=
\begin{cases}
N_{mol,sil}^{AGB} \, \mu_X \, N_{ato}^X \; \; \mbox{for} \; \frac{M_{ej,C}^{AGB}}{M_{ej,O}^{AGB}} <  0.75  \\
0 \; \; \mbox{otherwise}
\end{cases}
\end{equation}
It is straightforward to adapt the above treatment to chemical compounds different from olivine. 

\subsubsection{SNae II and Ia}
Unlike AGB winds, it is reasonable to assume that the explosive outflows produced by SNae are mixed only at a macroscopic level \citep[e.g.][and references therein]{dwek98}, which implies that these channels can produce both carbon and silicate dust at the same time. Hence:
\begin{equation}
M_{dust,C}^{SNx}=
\delta_{SNx} M_{ej,C}^{SNx} 
\end{equation}
\begin{equation}
M_{dust,X}^{SNx}=
N_{mol,sil}^{SNx} \, \mu_X \, N_{ato}^X 
\end{equation}
\begin{equation}
N_{mol,sil}^{SNx} = \delta_{SNx,sil} \min\limits_{X  \in \{Mg,Fe,Si,O\}} \left(\frac{M_{ej,X}^{SNx}}{\mu_X \, N_{ato}^X}\right)
\end{equation}
where ${SNx}$ stands either for ${SNII}$ or ${SNIa}$, \rev{and $M_{ej,X}^{SNx}$ is the  mass of the generic element $X$ ejected by a star particle through the ${SNx}$ enrichment channel during a given time-step.} In the case of SNae we decrease the dust condensation efficiency to $\delta_{SNII,C} = \delta_{SNIa,C} = 0.5$ and $\delta_{SNII,sil}=\delta_{SNIa,sil}=0.8$. These values are the same as those adopted in \cite{dwek98} and are meant to account for the grains destroyed by the SN shock or incomplete condensation of available elements.

\subsubsection{Shattering} \label{sec:shattering}
In the diffuse gas, large grains tend to have high velocities (typically $v = 10\mbox{ km s}^{-1}$, \citealt{yan04}) due to decoupling from small-scale turbulent motions \citep{hirashita09}. Therefore in such environments, large grains are most prone to shattering due to collision, forming small grains. In our framework, shattering is the process which originates small grains, without directly affecting the total dust mass. The growth of small grain mass occurs by accretion (Section \ref{sec:acc_cou}). 
The shattering timescale is derived following the prescription in Appendix B of \cite{aoyama17}. The gas environment diffuse enough to promote efficient shattering is identified by the condition $n_{gas} < 1\mbox{ cm}^{-3}$ \citep[see][]{hirashita09}, in which case:
\begin{align}
\tau_{sh} = 
\tau_{sh,0}  \left(\dfrac{0.01}{D_L} \right) \left(\dfrac{1 \, \mbox{cm}^{-3}}{n_{gas}}\right) & \, \, \, \mbox{if} \, \,  n_{gas}<1\mbox{ cm}^{-3}  
\label{eq:taush}
\end{align}
\rev{where $D_L = M_{dust,L} / M_{gas}$ is the dust to gas ratio for large grains}. The value of the proportionality constant suggested by \cite{aoyama17} is $\tau_{sh,0} = 5.41 \times 10^7 \mbox{yr}$, obtained assuming the typical grain velocity dispersion quoted above $v = 10\mbox{ km s}^{-1}$, a grain size of $0.1 \mu$m, and a material density of grains 3 g cm$^{-3}$.


\subsubsection{Accretion and Coagulation in dense molecular gas}
\label{sec:acc_cou}
Accretion of gas metals onto dust grains as well as coagulation of grains are processes whose timescale is inversely proportional to the gas density, and become significant only in the densest regions of molecular clouds, at $n_H \gtrsim 10^2 - 10^3$ cm$^{-3}$ \citep[see ][equations 7 and 8]{hirashita14}. These high densities are unresolved in cosmological simulations, therefore  we have to resort to a sub-resolution prescription to estimate the fraction of gas $f_{dense}$ that is locally in this condition.
\cite{aoyama17} simply assumed a fixed fraction. Although for their simulated galaxy the results are weakly sensitive to the adopted value, this solution could be in general unsatisfactory. In fact, $f_{dense}$ should depend on the local condition of the gas and on the numerical resolution reached in a simulation. We thus introduce a more flexible approach.  
According to the results of high resolution ($\lesssim 10$ pc) simulations \citep[e.g.][]{wada07,tasker09}  the probability distribution functions  of ISM density is well described by a log-normal function, characterized by a dispersion $\sigma$, found to lie in the range 2 to 3,  and a number density normalization parameter $n_0$.  
\begin{equation}
f_{pd}(n) \, dn = \frac{1}{\sqrt{2\pi} \sigma}\exp\left[ -\frac{\ln(n/n_0)^2 }{2\sigma^2}\right] d \ln n
\label{eq:pdf}
\end{equation}
Therefore  the mass fraction $f(>n_{th})$ of ISM gas above a given density threshold $n_{th}$ is  
\citep[][equation 19]{wada07}
\begin{equation}
f(>n_{th}) = \frac{1}{2}\left[1 - \erf\left(\left[\ln\left(\frac{n_{th}}{n_0}\right) - \sigma^2\right] \left(\sqrt[]{2} \sigma\right)^{-1}\right)\right]
\label{eq:fdense}
\end{equation}
We use this equation in order to estimate $f_{dense}\equiv f(>n_{th})$ for each SPH gas particle, we adopt $\sigma=2.5$ and $n_{th}=10^3$ cm$^{-3}$ in our fiducial model.  In order to set the local value of the density normalization $n_0$, we take into account that for the distribution given by equation \ref{eq:pdf}, the mass averaged density is  $\langle n \rangle_M=n_0 \exp(2\sigma^2)$ and we identify $\langle n \rangle_M$ with a suitable number density associated to the SPH particle. Since conditions suitable for accretion and coagulation can only be achieved in star forming multiphase particles (see Section \ref{sec:numerical}), we set $\langle n \rangle_M$ equal to the density of the cold phase $n_{cold}$. In the multiphase models of SF by \cite{springel03} $n_{cold}$ is computed from the condition of pressure equilibrium between the two phases, and assuming $T_{cold}=1000$ K, while the temperature of the ionized hot phase $T_{hot}$ is the SPH gas temperature. \rev{In our simulations, both the median and the standard deviation of $n_{cold}$ distribution increases with redshift, and range (in cm$^{-3}$) from about $\sim 1$ to $\sim 20$ and from a few tens to a few hundreds respectively.}

As already remarked, in the two size approximations, only small grains increase their mass when gas metals deposit on their surface. Following \cite{aoyama17}, we use the following expression to estimate the corresponding timescale: 
\begin{equation}
\tau_{acc} =  
\tau_{acc,0} \left(\dfrac{Z_{\odot}}{Z_{gas}} \right) f_{dense}^{-1} f_{cld}^{-1}
\label{eq:tauac}
\end{equation}
where $f_{cld}$ is the fraction of cold gas in multiphase particles (MP). Indeed
this equation does not apply to single-phase particles, nor to the hot phase of multiphase particles, since their density is far too low for the process to occur. \cite{aoyama17} set $\tau_{acc,0} = 1.2 \times 10^6 \mbox{yr}$.

In MP particles, small grains can also coagulate to form large grains. The timescale is given by \citep[see][]{aoyama17}
\begin{equation}
\tau_{co} =  
\tau_{co,0}
\, \left(\dfrac{0.01}{D_S} \right) \left(\dfrac{0.1 \, \mbox{km s}^{-1}}{v_{co}}\right)  f_{dense}^{-1} f_{cld}^{-1}
\label{eq:tauco}
\end{equation}
where \rev{$D_S = M_{dust,S} / M_{gas}$ is the dust to gas ratio for small grains} and $v_{co}$ is the velocity dispersion of small grains, for which a value of 0.1 km s$^{-1}$ is adopted based on calculations from \citet{yan04}. Just like with $\tau_{acc}$, the equation has been modified to consider only the cold phase of multiphase particles. \cite{aoyama17} used $\tau_{co,0} = 2.71 \times 10^5 \, \mbox{yr}$, a value derived assuming a typical size of small grains of $0.005\, \mu$m, a material density of 3 g cm$^{-3}$ and $n_{gas} = 10^3 \, \mbox{cm}^{-3}$.



\begin{table*}[ht]
	\def\arraystretch{1.5}
    \centering
	\begin{tabular}{ l  c  c  c  c  c  c  c  }
    \hline
    { run name} & { I.C.} & $M^{200}_{z=0}$ [$10^{14} M_{\odot}$] & {production} & {$\tau_{sh,0}$ [Myr]} & {$\tau_{co,0}$ [Myr]} & {$\tau_{acc,0}$ [Myr]} & {$\tau_{sp,0}$ [Myr]} \\ \hline 
    {\tt fid} & {D2} & 5.41 & new & {$54.1$} & {$0.271$} & {$1.20$} & {$\rev{5.5}$} \\ 
	{\tt f-crsp} & `' & `' & `' &  off & off & off & {$\rev{5.5}$}\\ 
	{\tt f-nosp} & `' & `' & `' &  {$54.1$} & {$0.271$} & {$1.20$} & off\\
    {\tt f-sp.2} & `' & `' & `' & `' & `' & `' & {$\rev{27.5}$}\\ 
    {\tt f-snII} & `' & `' & SNII-only & `' & `' & `' & {$\rev{5.5}$}\\ 
	{\color{black}{\tt f-dw}} & `' & `' & Dwek & `' & `' & `' & `' \\
    {\tt f-D3} & {D3} & 6.80 & new & `' & `' & 1.20 & `' \\
    {\tt f-D6} & {D6} & 15.5 & `' & `' & `' & `' & `' \\ 
    {\tt f-D1} & {D1} & 17.5 & `' & `' & `' & `' & `' \\ \hline
	\end{tabular}
    \caption{List of the test runs discussed in this paper with their respective parameters. Column 1 represents the run name used in the text. Column 2 is the chosen initial condition region. Column 3 is the $z=0$ $M_{200}$ of the main cluster of the region. Column 4 is the dust production method from stars: with "new" we refer to the prescription presented in this paper, assuring that dust grains at production respect a given proportion of O, Mg, Fe and Si, namely that  of olivine MgFeSiO$_4$ in this work, "dwek" is the prescription 
proposed by Dwek (1998) (see \ref{dsynth} for details).  Finally with "SNII-only" we refer to runs in which stellar dust production is active only for the snII channel, as done by Aoyama et al.\ (2017). This is to test the relative importance of the other two channels SNIa and AGB used in our full implementation. Columns 5 to 9 are the normalization timescales defined in the text for each ISM evolution process (\ref{sec:dusfor}).}
    \label{tab:simtab}
\end{table*}
  


\subsubsection{SNae destruction}
Dust grains are efficiently destroyed by thermal and non-thermal sputtering in SNae shocks \citep[for a review see][]{mckee89}. To avoid confusion, hereafter we refer to destruction by SNae events as SNae destruction, and we reserve the term sputtering to what we describe in Section \ref{sec:spu}. To treat this process we follow \cite{aoyama17}. Let $N_{SN}$ be the number of supernovae exploding over a given integration timestep $\Delta t$. 
For simplicity, we do not distinguish by now between the effects of SNII and SNIa.
However the code can be trivially adapted to take into account different values of the relevant parameters for the two classes, since the respective $N_{SN,II}$ and $N_{SN,Ia}$ are already computed separately.
The timescale for the process in the timestep can be written as
\begin{equation}
\tau_{SN} =  \frac{\Delta t}{1 - \left(1 - \eta\right)^{N_{SN}}}
\end{equation}
where
\begin{equation}
\eta = \epsilon_{SN} \min \left( \frac{m_{SW}}{m_{g}}, 1 \right)
\end{equation}
In this expression $m_g$ is the gas mass of the SPH particle, $m_{SW}$ is the gas mass swept by a single SN event\footnote{In our relative low resolution simulation $m_{s}<<m_{g}$.}, and $\epsilon_{SN}$ is the efficiency of grain destruction in the shock, that we set to the same reference value 0.1 used by \cite{aoyama17}. The shocked gas mass has been estimated in \cite{mckee89} as:
\begin{equation}
m_{SW} = 6800 M_{\odot} \left( \frac{E_{SN}}{10^{51} \mbox{erg}} \right) \left( \frac{v_s}{100\mbox{ km s}^{-1}}\right)^{-2}
\end{equation}
where $E_{SN}$ is the energy from a single SN explosion and $v_s$ the shock velocity. We adopt the fiducial value of $E_{SN}=10^{51}$ erg. \cite{mckee87} give the for $v_s$ the following expression
\begin{equation}
v_s = 200\mbox{ km s}^{-1} \left( n_0 / 1 \mbox{cm}^{-3} \right)^{1/7} \left( E_{SN} / 10^{51} \mbox{erg}\right)^{1/14}
\end{equation}
However, since the number density of ambient gas for single SN blasts is not resolved, we simply set $v_s = 200$ km s$^{-1}$. As a result, $\eta = 170 \, M_{\odot}/M_g$ in our simulations.

\subsubsection{Thermal Sputtering}
\label{sec:spu}
Dust grains can be efficiently eroded by collisions with energetic ions. This process becomes important whenever the ambient gas is hot enough,  $T_g\gtrsim 10^6$ K,  in which case it is dubbed {\it thermal sputtering}. To describe it, we employ the analytical formula given by \cite{tsai95}. This is an accurate enough approximations of detailed calculations for both carbonaceous and  silicate grains, at least for gas temperature smaller than a few times $10^7$ K. Above this temperature the process efficiency tends to stall \citep{tielens94}. 
Taking this into account and combining equations (14) and (15) of \cite{tsai95} we get:
\begin{equation}
\tau_{sp} = \tau_{sp,0}  \, \left(\frac{a}{0.1 \, \mu\mbox{m}}\right) \left(\frac{0.01 \, \mbox{cm}^{-3}}{n_g}\right) \left[ \left(\frac{T_{sp,0}}{\max(T_g,3\times 10^7\mbox{K})}\right)^\omega+1 \right]
\label{eq:tausp}
\end{equation}
with $T_{sp,0} = 2\times 10^6 K$, $\omega=2.5$ and \rev{$\tau_{sp,0} = 5.5 \times 10^6 \mbox{yrs}$}.
In this equation $n_g = \rho/\mu m_p $ is the number density including both ions and electrons. \rev{To derive the normalization constant $\tau_{sp,0}$, we adopted the mean molecular weight $\mu=0.59$ of a mixture of 75\% H and 25\% He, both fully ionized. Moreover, we have taken into account that for spherical grains the mass variations timescale is related to that for radius variations by $m / \dot m = a / 3 \dot a$. }
We adopt $a=0.05 \, \mu$m  and a ten times smaller value for the typical radii of large and small grains, respectively. The former is the average radius for a power law size distribution with index $-3.5$, ranging from 0.03 $\mu$m (the adopted boundary between small and large grains) to $0.25$ $\mu$m \citep[e.g.][]{silva98}.

\subsection{Fiducial run and its variations}

In \tab\ref{tab:simtab} we list a selection of interesting runs discussed in the paper. Most of the analysis refers to the "fiducial" run {\tt fid}, for which we essentially adopted the same parameters values as \citet{aoyama17}. These authors simulated an isolated spiral galaxy, and found results broadly consistent with spatially resolved observations of nearby dusty galaxies. It is however worth pointing out that we are working at resolutions lower by about three orders of magnitude in gas particle mass. Moreover, our model has two significant differences with respect to their work: we treat separately the production and evolution of the two distinct chemical species of carbon and silicate grains, and we include a treatment of thermal sputtering in the hot ICM. The latter process is crucial for galaxy cluster environments. Nevertheless, we believe that the parameter values defined by \citet{aoyama17} are a good starting point for our work. We will not explore the full dependence of dust properties on the parameters. Such work is postponed to future papers. As for sputtering, we adopt in the fiducial run parameter value derived from the classic work by \citet{tielens94}. 

In order to separate the other dust evolutionary processes from dust production and sputtering, we performed runs {\tt f-crsp}, where only the latter processes are turned on, and {\tt f-nosp} in which all processes but sputtering and SN destruction are active. To compare our dust production method, conserving the element fraction of olivine, to that by \cite{dwek98} we run {\tt f-dw} in which his method is instead employed. We also performed a run {\tt f-snII} where only SNII contributes to dust production, but not SNIa nor AGBs. With this run we test, and to some extent disprove, the claim by \cite{aoyama17} that dust production by SNII is sufficient to predict the dust properties. In {\tt f-sp.2} we increase the sputtering timescale by a factor 5, for illustrative reasons discussed in Section \ref{sec:obs}.  

These different runs were performed on the small mass region D2 of our set of initial conditions. The fiducial model was also applied to the other three regions D3, D1 and D6 (See Section \ref{sec:simclus}). 
Unless otherwise specified, the analysis that follows applies to the fiducial run {\tt fid}.


\section{Results} \label{sec:results}

In this section we study the behavior of the dust model within the simulation. First we follow the evolution of a couple token SPH gas particles representative of two interesting environments. A single gas particle could be thought of as the ISM of a one zone-model, such as those put forward by \cite{dwek98} and \cite{hirashita15}. 
Then we investigate the evolution of global properties of dust in the main cluster region. We conclude with a preliminary comparison to observations.

\subsection{Inside individual gas particles}
\label{sec:dustpart}

\begin{figure}
\centering
\includegraphics[width = \columnwidth]%
{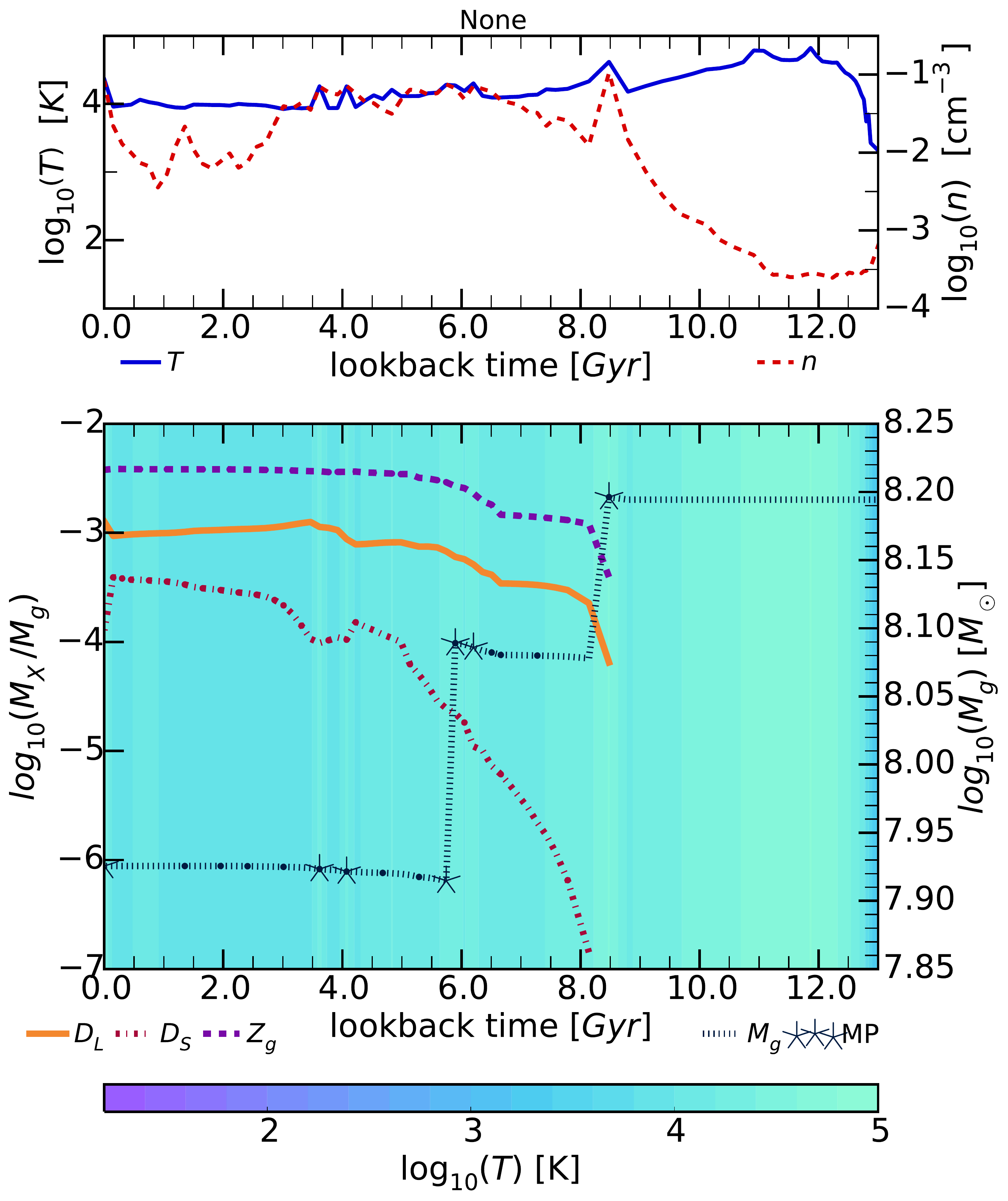}
\caption{For a gas particle in the fiducial run residing in a small, quiet, peripheral galaxy by $z=0$: Top and bottom plots represent the time evolution (\emph{x-axis} $t_{lb}$ time in Gyrs, with 0 being today) of various gas properties: (\emph{\bf top plot}) temperature (\emph{left y-axis, blue solid line}) and number density (\emph{right y-axis, red dashed line}). (\emph{\bf bottom plot}) Gas metallicity (\emph{left y-axis, dashed purple line}), as well as the Dust-to-Gas (DtG) for  large dust grains (\emph{left y-axis, solid orange line}) and small dust grains (\emph{left y-axis, dot-dashed red line}), total mass (\emph{right y-axis, dotted dark blue line}) of the gas particle. The stars mark the snapshots in which the gas particle was captured in multiphase state (MP).}
\label{fig:partisolated}
\end{figure}
 
\begin{figure}
\centering
\includegraphics[width = \columnwidth]{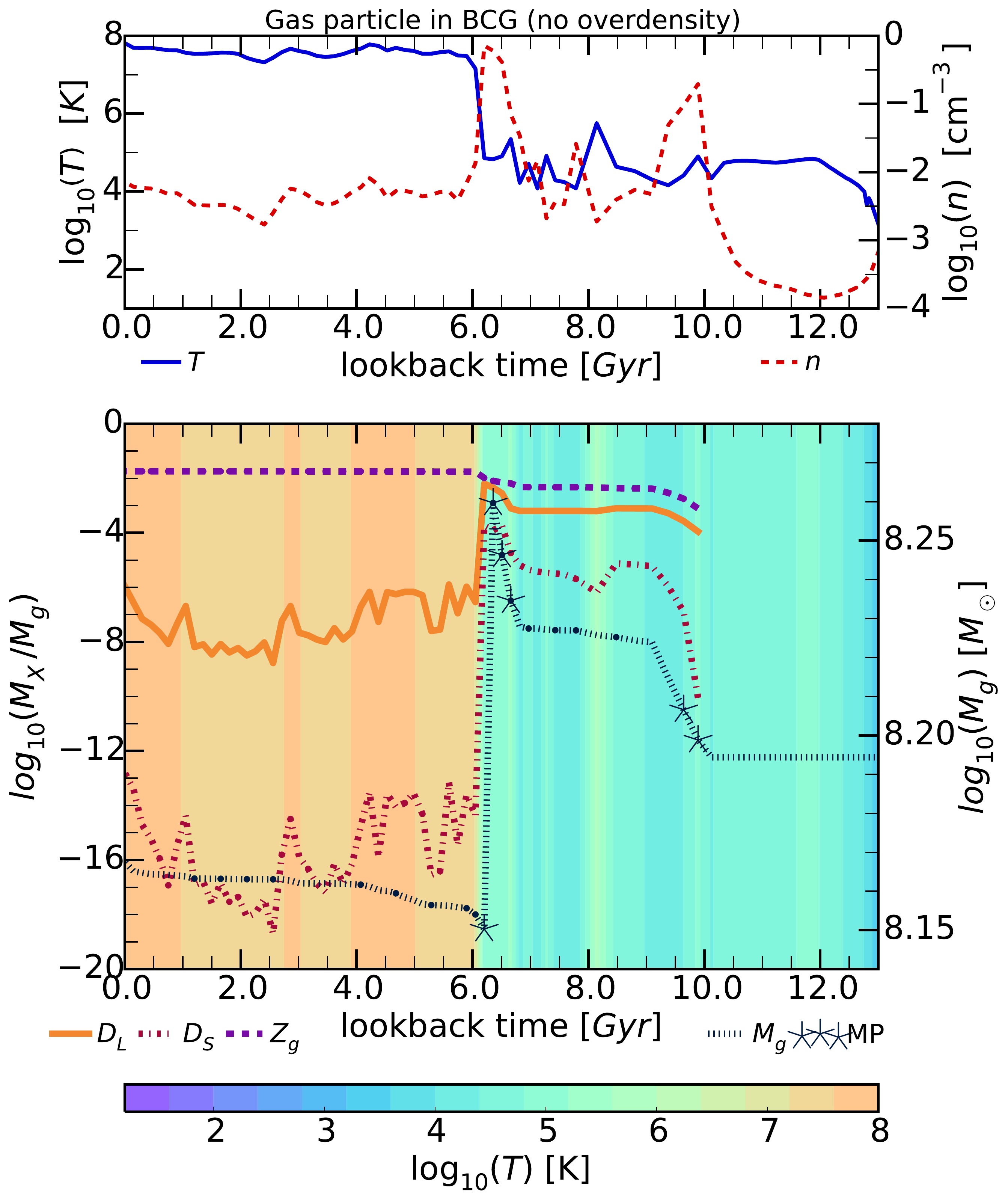}

\caption{Same as \fig\ref{fig:partisolated} but for a gas particle residing in the BCG by $z=0$. The colorbar gradient for temperature is fixed for both of the lower plots in the two figures.
}
\label{fig:partprocesses}
\end{figure}

In \fig\ref{fig:partisolated} we follow the dust, gas metal, total mass, temperature, and density of a simulated gas particle which by $z=0$ is located within an isolated galaxy at the periphery of a cluster in our fiducial run. This particle features a relatively quiet evolution. It spawns two star particles, around a lookback time ($t_{lb}$) of 8.5 Gyr and 6 Gyr, as a consequence of its star forming state. The particle's mass drops to about 75\% and 50\% of its original value during each episode (see Section \ref{sec:numerical}; dotted dark blue line  of the bottom panel). 
It begins to be enriched in gas metals and large grains by the neighboring stellar particles at a $t_{lb}$ of 8.5 Gyr,  around the time of star formation. 
Soon after that, shattering begins to produce a population of small grains. Its growth rate is steady for the first 4 Gyrs, thanks to a fairly constant gas density (red dashed line in the top panel). The gas temperature does not deviate significantly from $10^4$ K throughout its history, whether the gas density starts low at $3 \times 10^{-3}$ cm$^{-3}$ and then fluctuates by 1 dex below $10^{-1}$ cm$^{-3}$. The second star particle formation episode at 6 Gyrs does not affect the DtG and metallicity, as these quantities decrease proportionally to the mass loss.

While the particle is multiphase, accretion sticks gas metals onto small grains and small grain coagulation in turn convert small grains to large ones. The seven multiphase snapshots are highlighted by a star symbol on the mass line of the bottom panel.
The last three ones, two around 4 Gyr in $t_{lb}$ and the third very recent, occurs while the small grain abundance is within 1 dex to that of large grains. In this case, coagulation prevails over accretion, and therefore a depletion of small grains is visible, whilst a tiny bump in the large grains mass ratio can be recognized. Note that it is possible that the particle underwent other multiphase periods between two subsequent snapshots. Similar periods are not highlighted by star symbols in the plot. 

 \fig\ref{fig:partprocesses} represents the evolution of another particle which instead ends up in the brightest cluster galaxy (BCG) by $z = 0$. 
This gas particle is first enriched with gas metals and large grains as soon as it becomes multiphase and enters a cold overdensity of star forming gas. This period lasts a few hundred million years around 10 Gyrs in $t_{lb}$. The particle density spikes, and with it shattering increases and produces small grains. 
At 8 Gyr the particle temperature increases above $10^5$ K for a while, increasing the efficiency of sputtering enough to cause a small dip of DtG, visible (almost) only for small grains. This is because the sputtering timescale for small grains is 10 times smaller than that for large grains. At a $t_{lb}$ of about 6.5 Gyrs the particle enters another cold overdensity. At this time the large grains abundance approaches values, similar to the expected ISM values in galaxies \citep{li01IR}, close to the gas metal abundance via coagulation, and also small grains manage to rise via accretion.  Shortly after, at 6 Gyrs in $t_{lb}$, this gas particle spawns a star particle. Soon after it enters in a hot state $T\gtrsim 10^7$ K, where it remains until the present time.
In the hot state, sputtering becomes a strong source of dust destruction. Notice again the steeper decay  of small grains caused by the size dependence of sputtering. The fluctuations in dust abundances after 8 Gyrs are due to slight enrichments from the surrounding star particles. While these contributions are negligible in the relatively high gas metals mass fraction, they are evident in that of large grains, and even more so in that of small grains. 

\subsection{Evolution of the global properties of dust}


\begin{figure*}
	\centering
	\includegraphics[width=.98\textwidth]{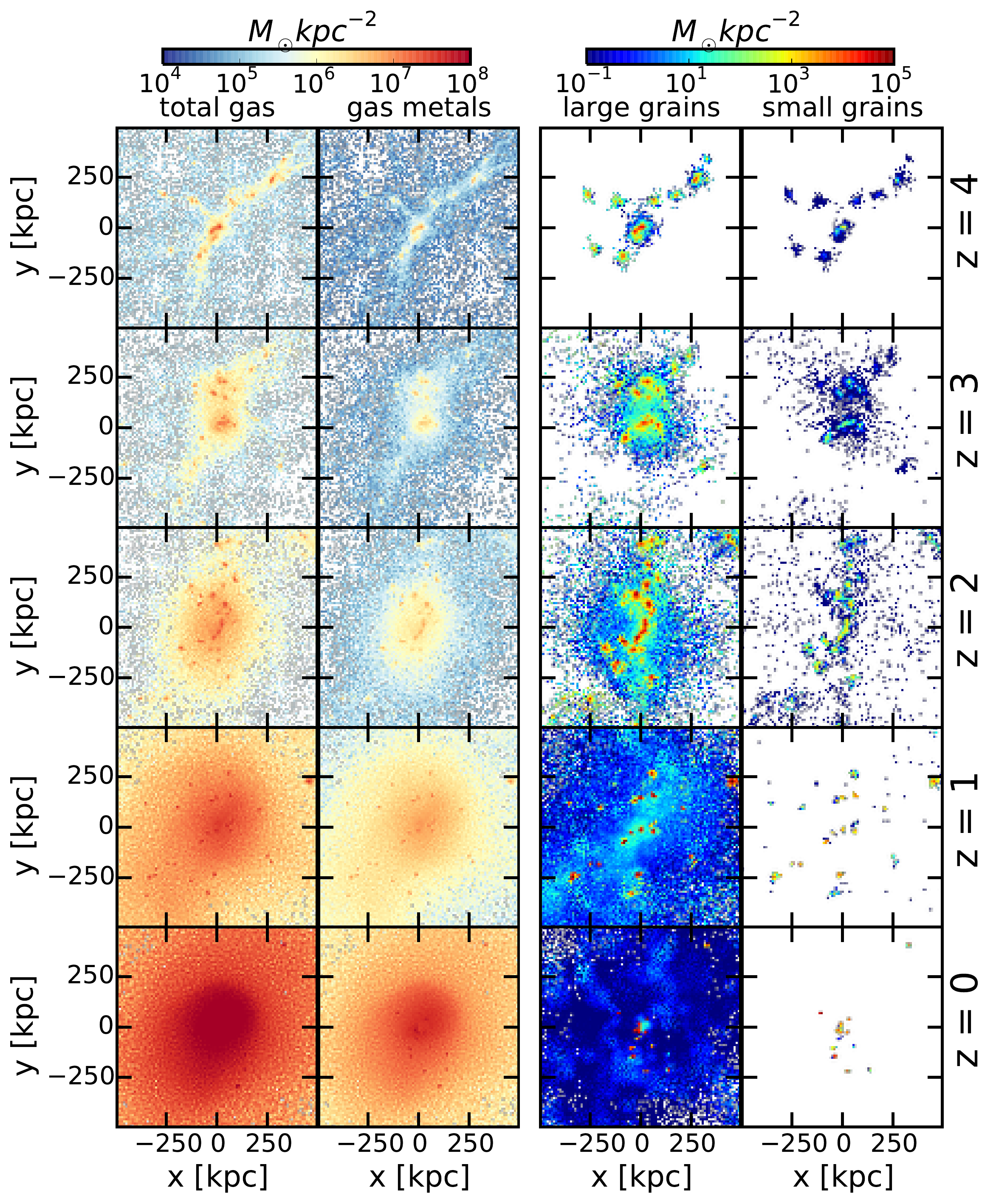}
	\caption{For the {\tt fid} run, column density maps for total gas mass, gas-phase metals, large dust grains, and small dust grains, in a box of 1 Mpc in physical size over 5 redshifts (from top to bottom, z = 4, 3, 2, 1, and 0). The two colorbars are fixed at all redshifts from total gas and gas metals and for large and small grains respectively. Dust abundances trace gas mass distributions until about $z=2$. After $z=2$ sputtering destroys dust. Small grains evolve to the point of reaching large grains abundances only in cold overdensities, but they are destroyed more efficiently than large grains in hot gas particles.}
    \label{fig:distrclus}
\end{figure*}

\begin{figure*}
	\centering
	\includegraphics[width=.99\textwidth]{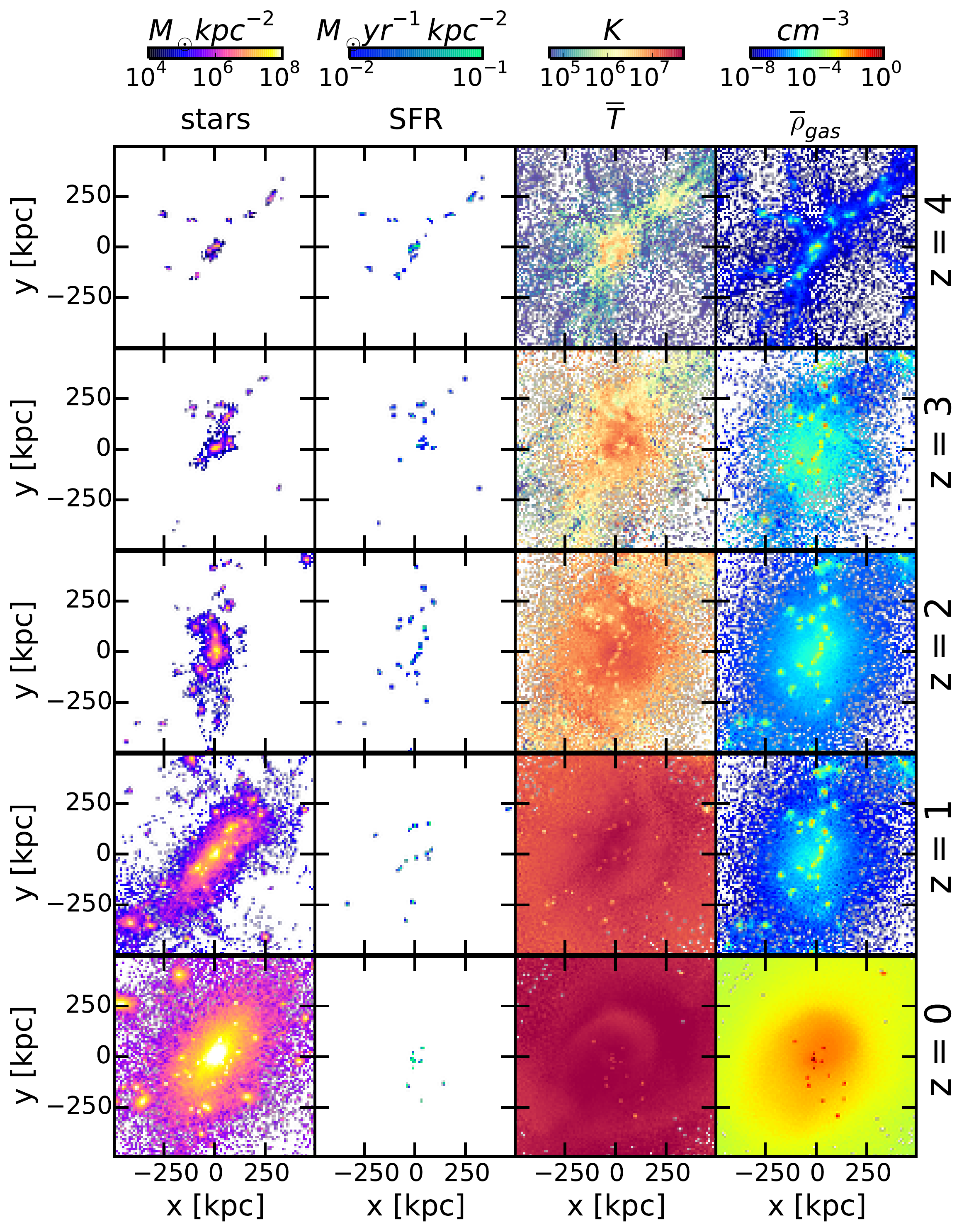}
	\caption{Similarly to \ref{fig:distrclus} for the {\tt fid} run, columns represent from left to right: maps of stellar mass and SFR column densities, and of mean mass-weighted temperature and mean number density in a box of 1 Mpc in physical size over 5 redshifts (z = 4, 3, 2, 1, and 0). Both means are weighted with the gas particle masses. Star formation occurs in cold overdensities mostly occupied by multiphase gas particles.}
    \label{fig:distrclusprop}
\end{figure*}


\subsubsection{Dust distribution in the cluster region} \label{sec:dustdistr}

In this section we analyze the global evolution of dust distribution around the central cluster. \fig\ref{fig:distrclus} follows from top to bottom the redshifts $z=4, 3, 2, 1, 0$, and from left to right the column densities of total gas mass, gas-phase metals, large dust and small dust grains. At each redshift we project a physical cube of size 1 Mpc around the main cluster's progenitor. Regions in which the column density is lower than $10^{-8} \sigma_{peak}$ where $\sigma_{peak}$ is the peak column density, are omitted. 

The maps shows that gas-phase metals approximately mirror the total gas distribution at all redshifts. Large grains follow loosely a similar pattern until $z \sim 2$. At lower z, this correlation breaks down since many particles reach  $T \gtrsim T_{sp,0} = 2\times 10^6$ K, above which sputtering erodes grains effectively, particularly small ones. 
Moreover, small grains are less abundant than large ones at all times and do not grow as rapidly, except in cold ($T < T_{MPh} = 5\times 10^5$ K) star forming gas over-densities, where shattering and accretion work efficiently. 



\fig\ref{fig:distrclusprop} refers to the same 1 Mpc physical cube around the main progenitor. It represents from left to right the star particle column density, the star formation rate (SFR) column density, the mass weighted mean  temperature and mean number density of gas particles. These maps help in interpreting the former ones. 
For instance, it can be appreciated that the  structures wherein dust survives and evolves undisturbed by sputtering are characterized by low temperature. 
They feature SFR between 0.01 and 0.1 $M_{\odot} yr^{-1} \mbox{kpc}^{-2}$. As the cluster gains mass its average temperature rises, and sputtering begins to destroy efficiently dust in the ICM.

\begin{figure}
	\centering
	\includegraphics[width = \columnwidth]{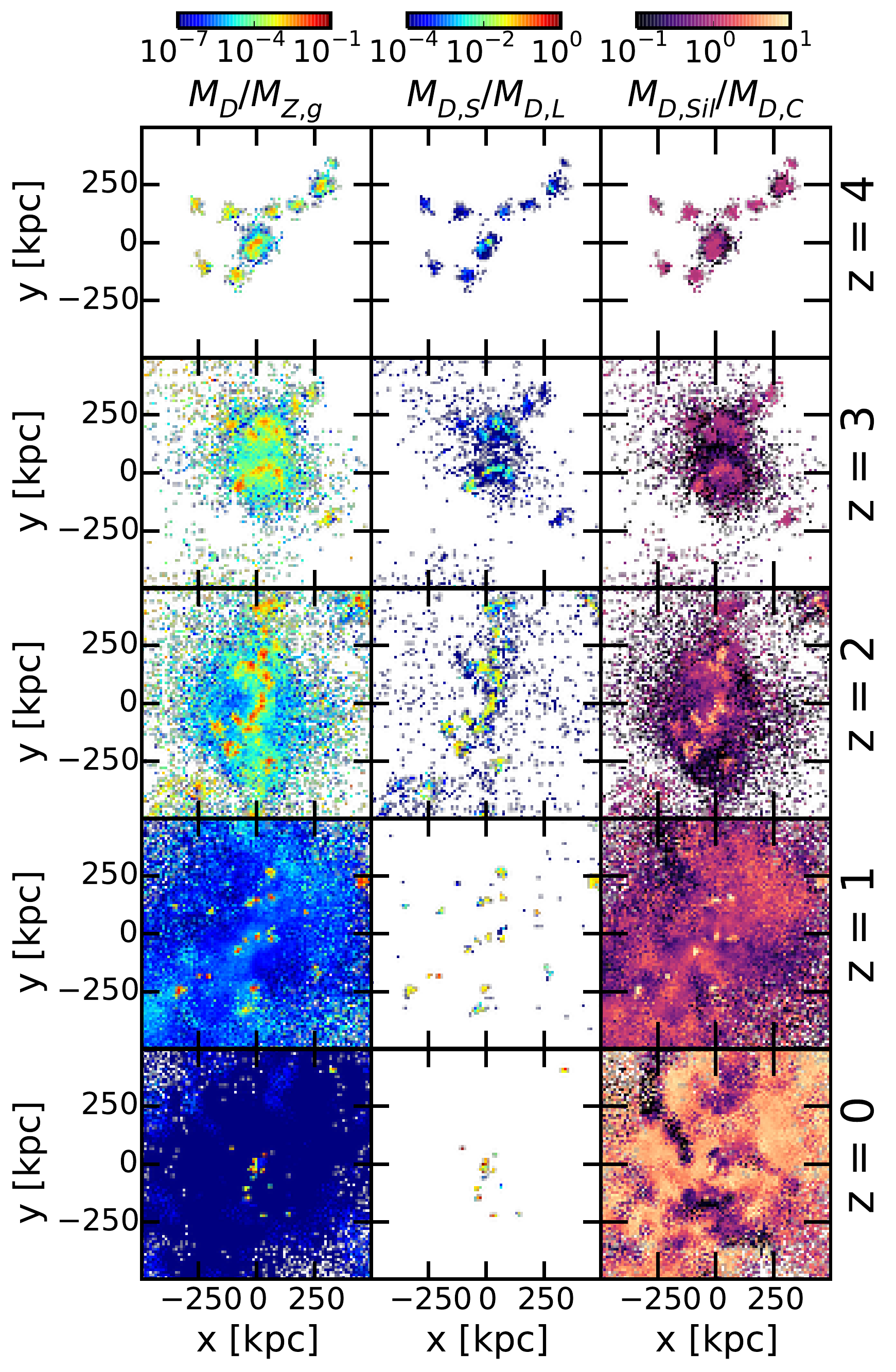}
	\caption{Similarly to \fig\ref{fig:distrclus}, columns represent maps of the dust-to-gas-metal ratio (left), small-to-large grain ratio (center) and silicates vs carbonaceous dust (right) in a box of 1 Mpc in physical size over 5 redshifts (z = 4, 3, 2, 1, and 0). }
    \label{fig:distrclusdtz}
\end{figure}

\begin{figure}
	\centering
	\includegraphics[width = \columnwidth]{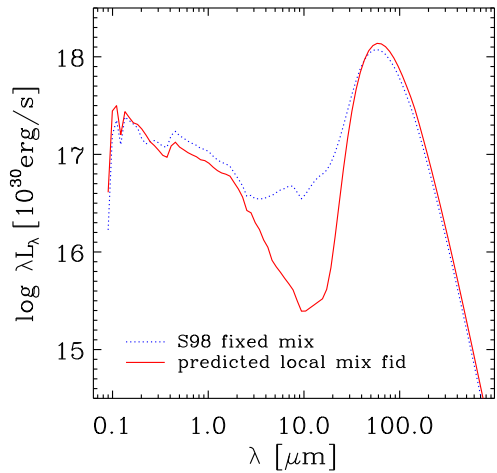}
	\caption{Spectral energy distributions predicted by SKIRT \citep{camps15} for the 100 kpc central box of the region D2 at z=4. The dotted blue line has been computed adopting for the whole volume a "standard" dust mixture reproducing the properties of dust in the diffuse ISM of the MW. The solid red line instead takes into account the point to point variations predicted by the simulation for the relative abundances of small and large grains, as well as those for silicate and graphite grains. See Section \ref{sec:dustdistr} for more details.} 
    \label{fig:skirtseds}
\end{figure}

\begin{figure}
	\centering
	\includegraphics[width = \columnwidth]{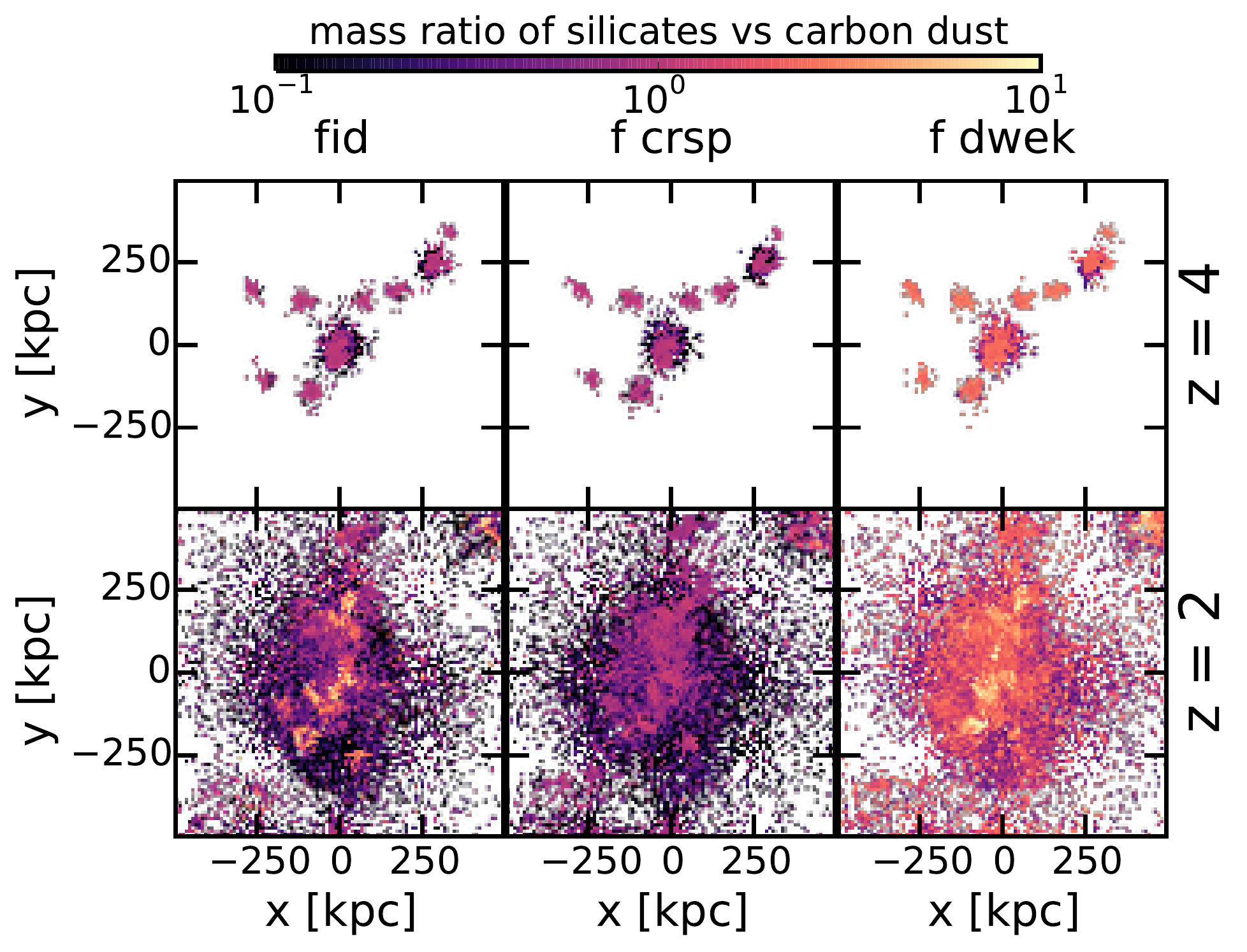}
	\caption{Similarly to \fig\ref{fig:distrclusdtz} right column,
The three columns compare maps of the ratio for the fiducial run ({\tt fid}), the run including only dust production from stars and sputtering({\tt f-crsp}), and the run in which the  silicate dust is produced by stars adopting the prescription by \citet{dwek98} 
}
    \label{fig:distrsilc}
\end{figure}

\fig\ref{fig:distrclusdtz} maps the average dust-to-gas-metal ratio (left), small-to-large grain ratio (center) and silicates vs carbonaceous dust (right). At $z>3$, when the SF activity in the proto-cluster region is at its maximum and dust reprocessing is expected to be important, dust properties are predicted to differ significantly from those derived for the MW dust. The latter are almost always adopted in computations to account for dust reprocessing \citep[e.g.][and references therein]{dominguez14}. For instance, compared to the dust models proposed for the MW by \citet{weingartner01}, in the central 100 kpc at z=4 the mass ratio of small/large carbon grains is more than 2 dex smaller, while the silicate/carbon mass ratio is about twice  smaller\footnote{For the most commonly adopted model 4 by \cite{weingartner01}, the mass ratio of small to large C grains is 0.34, while the mass ratio of Silicate to Carbon grains is 2.5. The model proposed by \cite{silva98} for the MW dust, usually adopted in GRASIL code features values not significantly different.}. We illustrate the possible consequences of these differences on the predicted SEDs in \fig\ref{fig:skirtseds}. Here we show two SEDs computed with the public radiative transport code SKIRT\footnote{http://www.skirt.ugent.be} \citep{camps15}. In one case we have simply used a dust mixture similar to that adopted so far in most GRASIL \citep{silva98} and all GRASIL3D \citep{dominguez14} applications\footnote{These codes are often employed to compute dust reprocessing in semi-analytic models and galaxy formation simulations respectively. They both  allow for custom variations of the mixture, albeit this feature is seldom used due to lack of information. A spacial dependence on these properties is not implemented in the present versions.}. Here, the only  information on dust derived from the simulation is the total dust content of each SPH particle. In another computation we have instead exploited  the full information concerning the relative partition in the four categories of dust grains (graphite and silicate, small and large), by locally adjusting the adopted mixture at the position of each SPH particle, in order to reflect this partition\footnote{\rev{We fed SKIRT with a superposition of four spatial distribution of dust densities, each one to represent one of the four grain types followed by our model. The size distributions for each of them, provided by suitable input parameters of SKIRT, were the same as that proposed by \cite{silva98}, but having 0.03 $\mu$m as the limiting size to distinguish between small and large grains respectively. Moreover, the normalization  of each of the four distributions has been computed at each point according to the local density of the corresponding grain type, as predicted by the simulation.}}. As it can be seen, the differences are important, particularly in the optical to mid-IR regime. We plan to explore in detail the observational consequences of these differences in the near future.

\begin{figure*}
\centering
\includegraphics[width = \columnwidth]{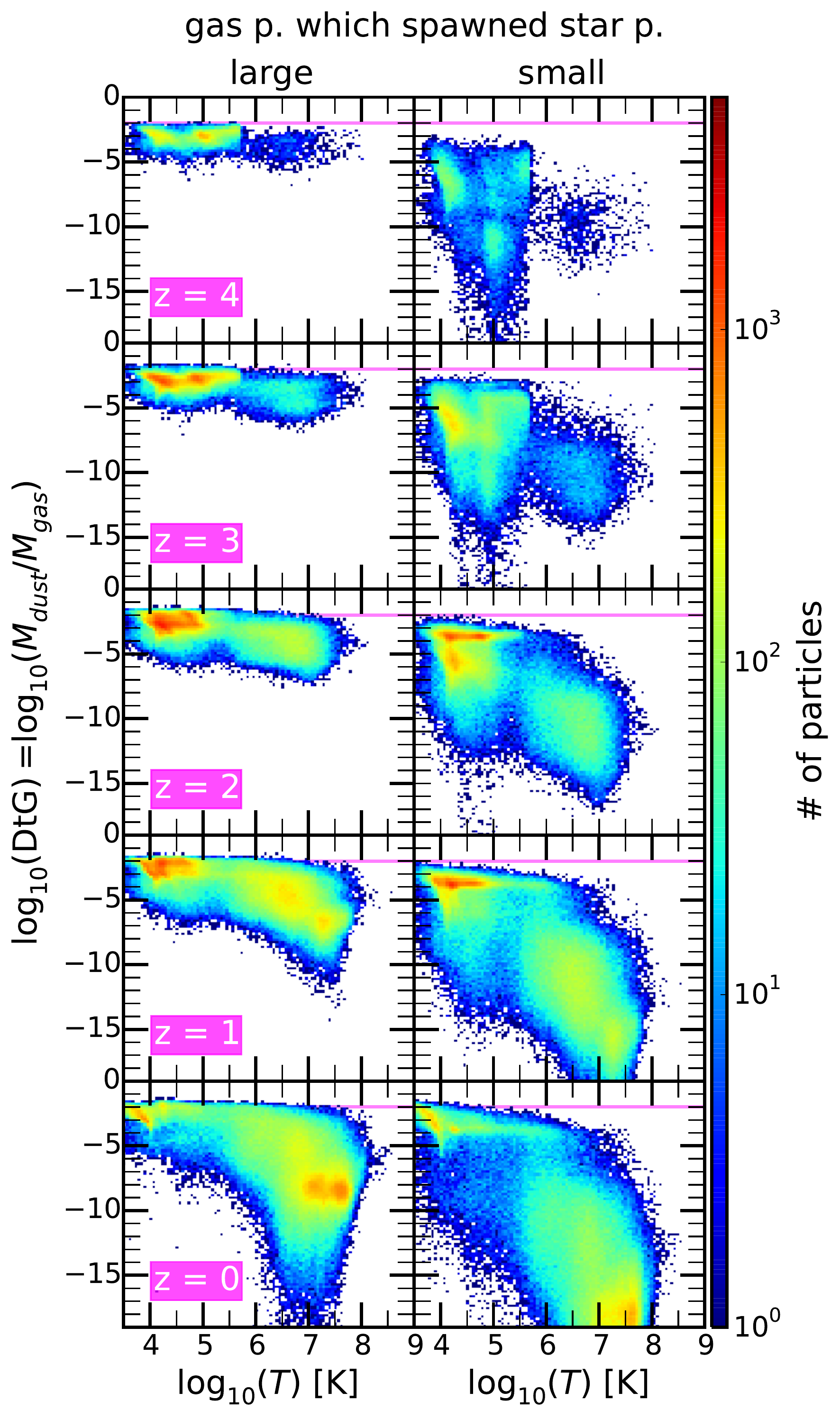}
\includegraphics[width = \columnwidth]{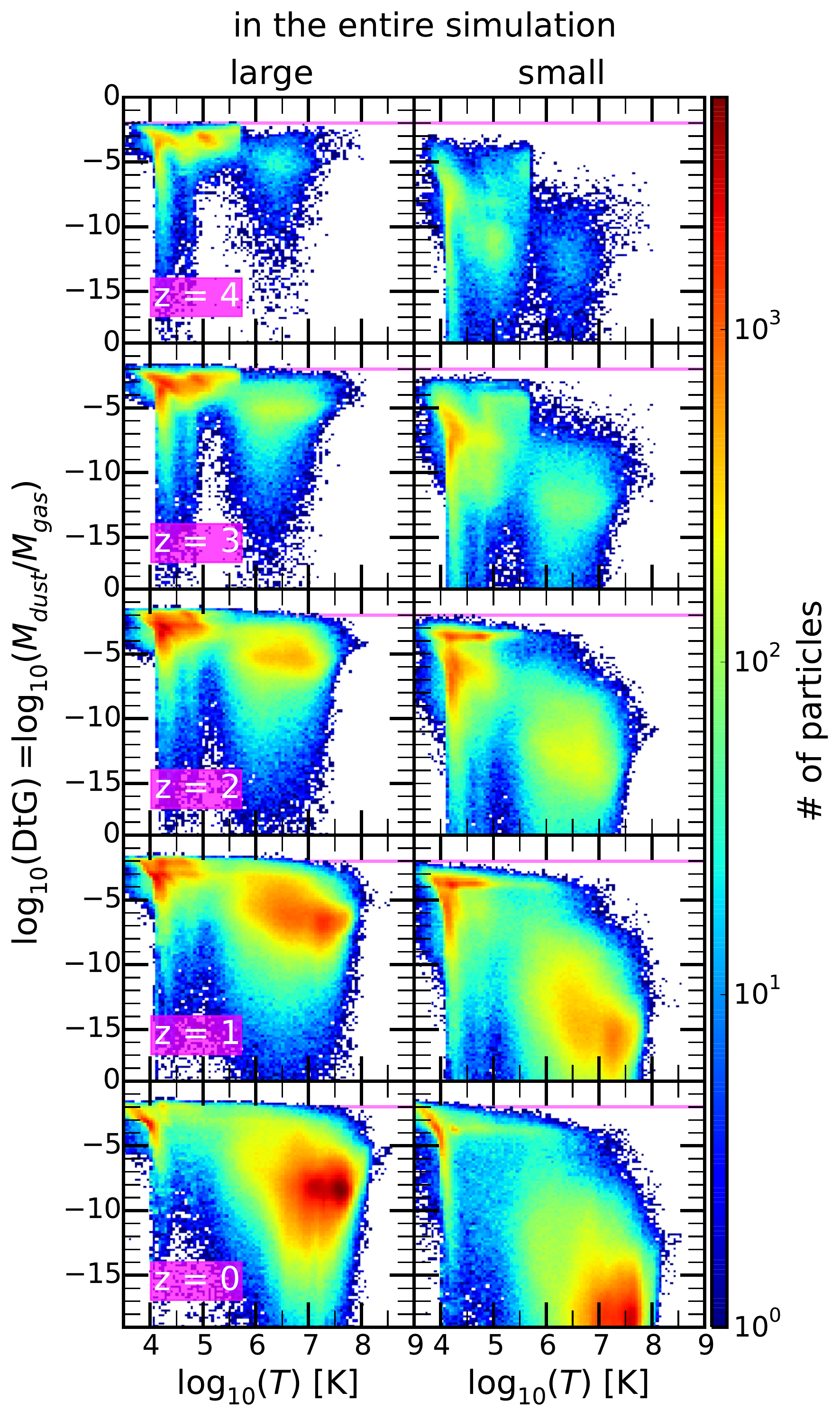}
\caption{2D Histograms of the total DtG vs temperature for gas particles which have spawned star particles in their past (left) and in the entire simulation (right) for the fiducial run and over 5 redshifts (z = 4, 3, 2, 1, and 0). The magenta line represents DtG =$10^{-2}$, close to the commonly accepted ISM dust abundance in the MW.}
\label{fig:dTscatterclus}
\end{figure*}

\begin{figure}
	\centering
	\includegraphics[width = \columnwidth]{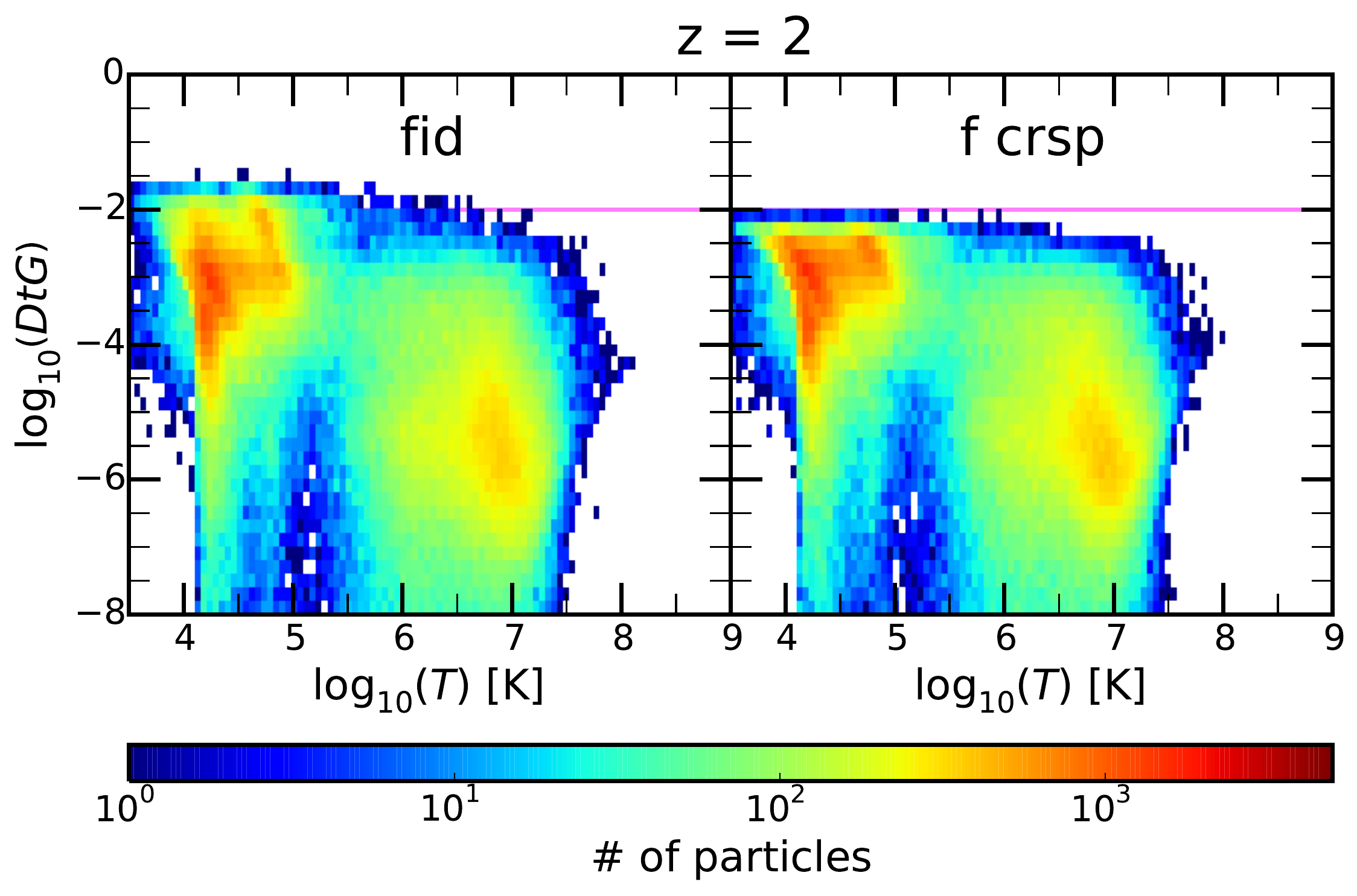}
   	\caption{
    2D histogram of DtG vs temperature within the {\tt fid} and {\tt f-crsp} at $z = 2$ including all gas particles.}
    \label{fig:dTscattersingle}
\end{figure}

The ratio silicate-to-carbon dust takes some time to reach values close to those adopted in "standard" dust mixtures. Actually the ratio arising from dust production by stars is significantly lower in the model, and its increase occurs via dust evolution in the ISM.
The behavior can be appreciated from the maps shown in Fig.\ref{fig:distrsilc}. From left to right, the three columns compare maps of the ratio for the fiducial run ({\tt fid}), the run including only dust production from stars and sputtering ({\tt f-crsp}), and the run in which the  silicate dust is produced by stars adopting the prescription by \citet{dwek98} ({\tt f-dw}, see Section \ref{dsynth}) respectively. At early time z=4, the first two runs are virtually indistinguishable. It is also worth noticing that the last one is characterized by substantially higher Sil-to-C ratios, already closer to, or even higher than, the standard one. Indeed the prescription put forward by \citet{dwek98} is substantially 
more liberal in using the ejecta to produce silicate dust. At lower redshift z=2, when evolution in the ISM has been important (if included in the computation), the fiducial run has increased the ratio in most of the region. This is not the case for the production and sputtering only run ({\tt f-crsp}, lacking all the processes causing dust evolution in the cold ISM). Also the {\tt f-dw} run does not show a sizable increment with respect to z=4, but in that case it was already high. This is because it leaves by construction less silicate elements in the gas phase at stellar production, meaning that gas accretion onto grains has less to add. The time evolution of the silicate-to-carbon ratio, integrated over $r_{200}$ is \rev{also shown in a later figure}.

\subsubsection{Temperature dependence of dust contents}

Additional insights on how and when the various processes affect the dust content of the SPH gas particles can be obtained from inspection of \fig\ref{fig:dTscatterclus}. Here we plot 2D histograms for DtG (y-axis) vs temperature (x-axis) over the usual 5 redshifts. DtGs are separately shown for large and small grains. The figure on the right refers to the entire simulation. That on the left includes instead only particles that have spawned at least one stellar particle before. This is meant to select preferentially gas particles which 
have spent a significant fraction of their life in a star forming environment. 

At $z \ge 3$ large grains  show a peak in the two dimensional distributions at $T \lesssim 10^{4}$ and $\mbox{DtG} \sim 10^{-4}$. This last value is about two orders of magnitude below the standard Galactic DtG of $\simeq 0.01$, marked by horizontal magenta lines in the panels. Indeed, at these early epochs the evolutionary processes in the ISM have not had sufficient time to affect much the dust content of most SPH particles. We have verified this by comparing with the run {\tt f-crsp}. The major effect of ISM evolution before $z \simeq 3$ is the production of a certain amount of small grains by sputtering. ISM evolution effects manifest appreciably at $z=2$. A population of gas particles featuring DtG close to $10^{-2}$ for large grains and $10^{-3}$ for small ones shows up  at $T \lesssim 10^5$. The particles belonging to this population have undergone multiphase periods, during which accretion onto small grains and their coagulation to form large ones have raised the DtG up to values similar to the Galactic one. Indeed these local peaks appears very similar in the right and in the left figures. As pointed out, the latter is meant to select gas particles characterized by multiphase star forming periods in their past. Moreover, in run {\tt f-crsp}, gas particles featuring DtG $ \ge 10^{-2}$ are not produced, as it can be seen in  \fig\ref{fig:dTscattersingle}. In this case the peak of the DtG distribution remains 1 to 2 orders of magnitude lower even at $z \le 2$, only becoming more and more populated.

At still lower redshift the high temperatures reached by most SPH gas particles promote efficient thermal sputtering. As a result, a well defined  maximum in the 2D histograms of the entire simulation plots develops at $T \gtrsim 10^7$
and at very low DtG$\lesssim 10^{-5}$ for large grains. The peak for small grains occurs at an even smaller DtG, since they are more strongly disrupted by sputtering. 

  \begin{figure}
  \centering
  \includegraphics[width = \columnwidth]{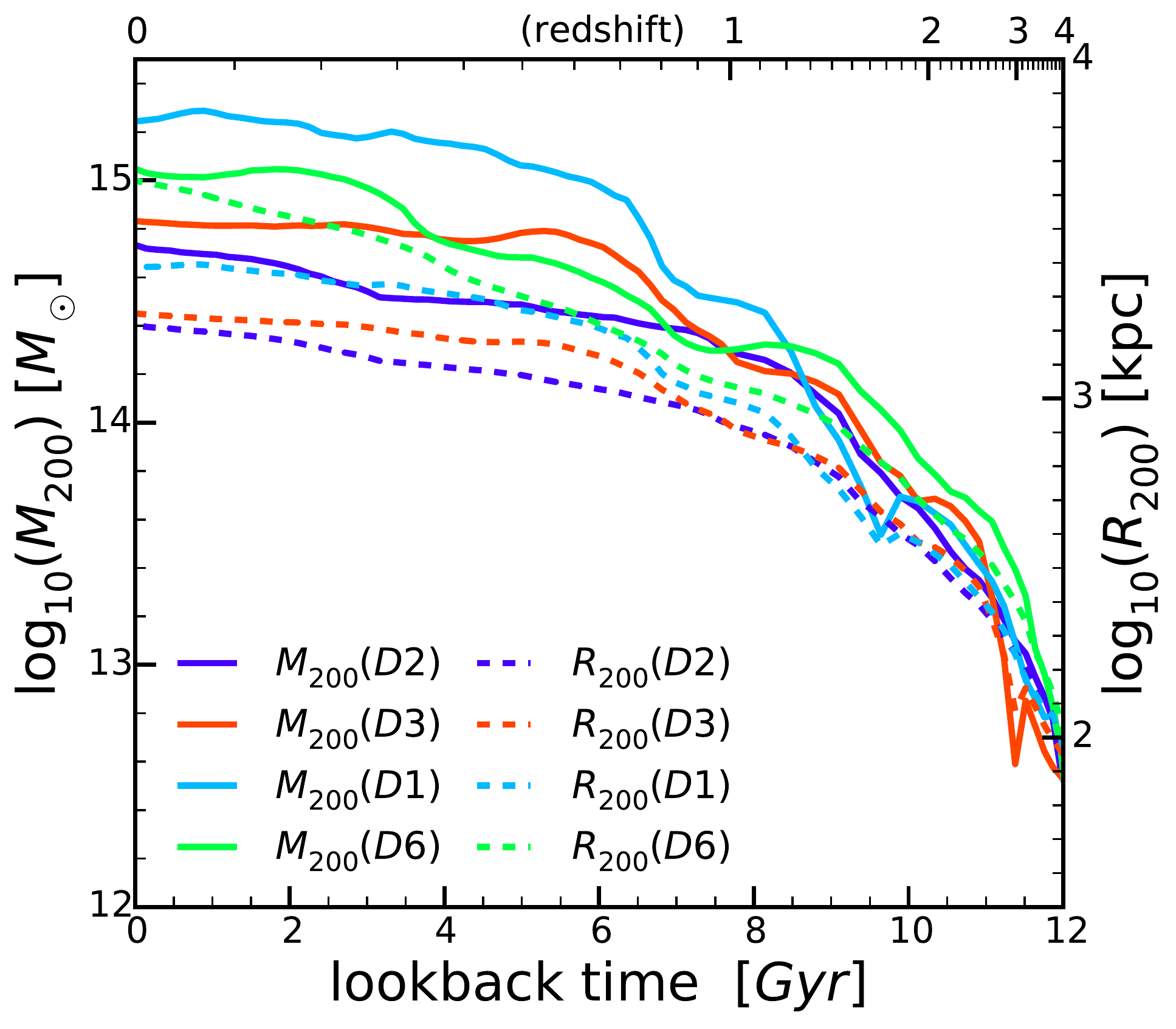}
  \caption{M200 and R200  of the fiducial run for the D2, D3, D1, and D6 regions.}
  \label{fig:M200massevol}
  \end{figure}
 
\begin{figure*}
\centering
\includegraphics[width = .9\textwidth]{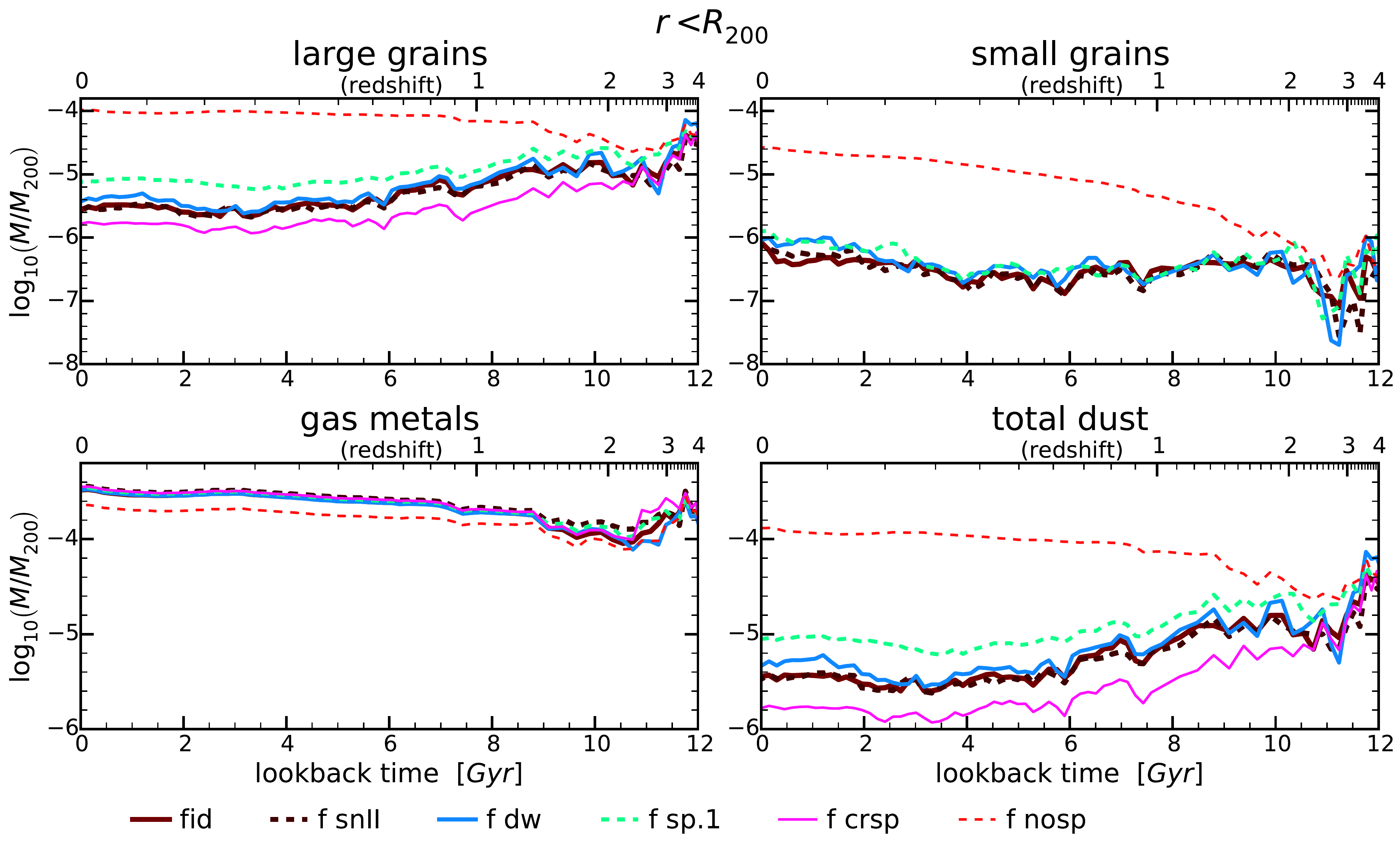}
\includegraphics[width = \textwidth]{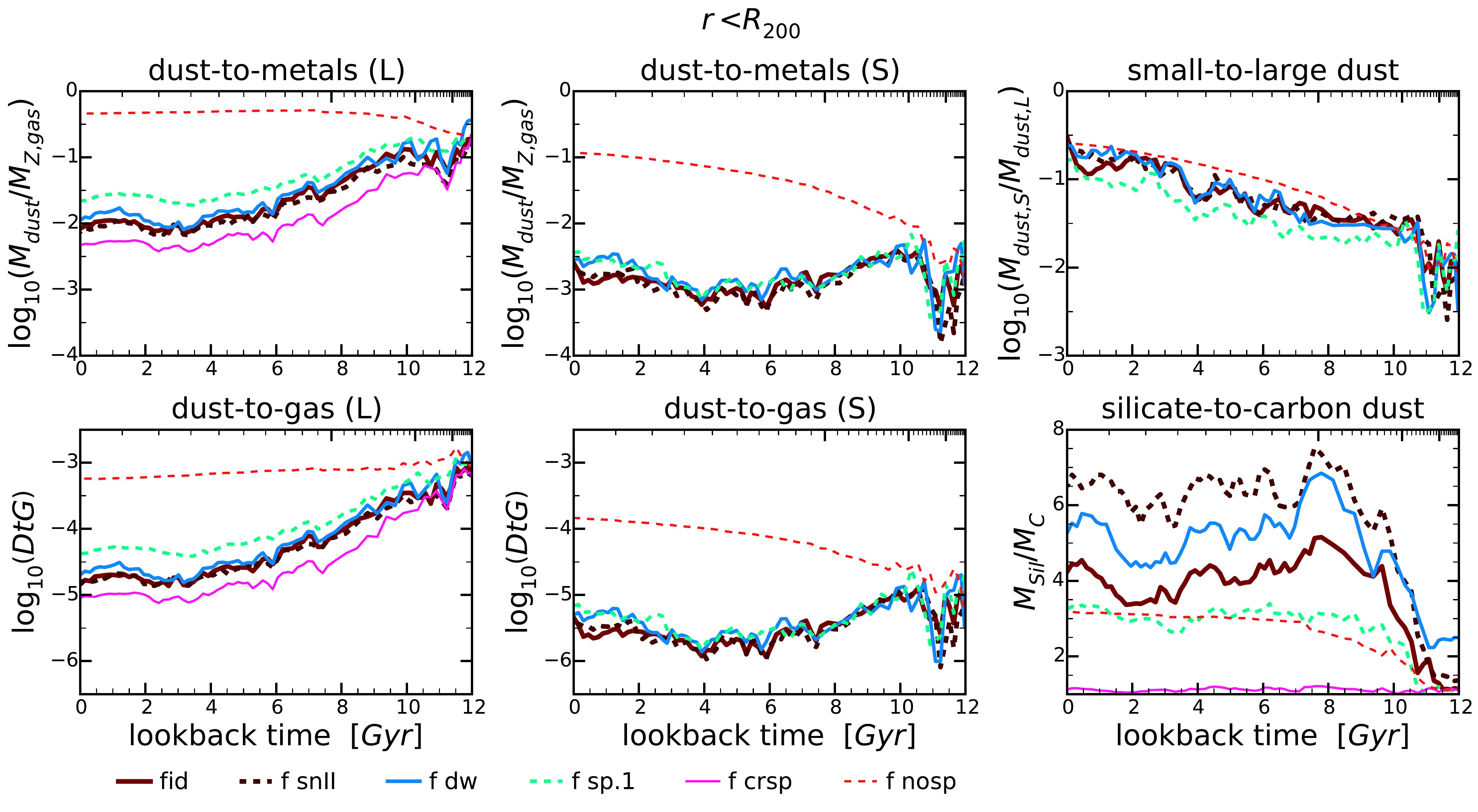}
\caption{Time evolution within $R_{200}$ of the main progenitor for a selection of runs. ({\textit{top 4 plots}}) masses of large grains, small grains, gas metals, total dust mass and ({\textit{bottom 6 plots}}) large-dust-to-gas-metal, small-dust-to-gas-metal, small-to-large ratios, DtG for large and small grains, and lastly the mass ratio between carbonaceous dust and silicates. Small grains are affected more strongly than large grains by the timescales of the evolution processes.}
\label{fig:allruns}
\end{figure*}

\subsubsection{Evolution history of run variations} \label{sec:evol}

\fig\ref{fig:allruns} illustrates the evolutionary history of various masses computed within $R_{200}$, in the main progenitor of the $z=0$ cluster. We include results for a selection of run variations on the D2 region. The 4 top plots show the gas metal and dust masses normalized to the evolving $M_{200}$, while the 6 plots in the bottom show various interesting ratios. In order to facilitate the interpretation, the evolution of $M_{200}$ and $R_{200}$ can be seen in  \fig\ref{fig:M200massevol}.

From the total dust panel we can appreciate that, excluding the run without sputtering {\tt f-nosp}, the dust production is faster than or comparable to the increase of the cluster mass only in the $t_{lb}$ lapse between 11 and 9 Gyr and at  $t_{lb} \lesssim 4$ Gyr. In the former interval, this is achieved thanks to the combined effect of shattering, accretion and coagulation occurring in the the gas, which enhance the dust content by a factor $\sim 2-3$ after $t_{lb} \sim 10$ Gyr, as can be understood by comparing the runs {\tt fid} with {\tt f-crsp}. In the latter run, the three above mentioned processes are switched off, and the ratio $M_{dust}/M_{200}$ is monotonically decreasing down to low $z$. The late $t_{lb} \lesssim 4$ Gyr flattening of the evolution is instead related to slow down of the $M_{200}$ increase and to the fact that sputtering has already destroyed most of the dust in the hot ICM.

By adopting the \citep{dwek98} recipe to compute the production rate of silicate grains from stars (run {\tt f-dw}), more liberal than our fiducial method imposing the chemical composition of olivine, we get about 50\% more dust at early time $t_{lb} \gtrsim 9$ Gyr, and a higher ratio $M_{sil}/M_{C}$ over the whole evolution.

It is interesting to note that by considering only the SNII channel for the production of dust (run {\tt f-snII}), the total dust content is somewhat under-predicted with respect to the fiducial run by up to $\sim 50\%$. But what is more important is that the ratio $M_{sil}/M_{C}$ is significantly over-predicted. Thus this approximation, sometimes adopted in other works \citep[e.g.][]{hou16,aoyama17,hou17,chen18} seems to be insufficient at least for certain purposes, such as computing the radiative effects of dust. 

As we already pointed out, in the fiducial run the mass fraction of small grains at high redshift, when SF is most active in the simulated region is much lower than that of standard mixtures ($\sim 0.2$). These mixtures are calibrated on the average properties of MW dust. This result is quite robust, in the sense that occurs in all the run variations we considered.
On the other hand, the ratio $M_{sil}/M_{C}$ is not very different from the standard value $\sim 2$ at early time, while it ends up significantly higher in most runs. The only exception are those runs with no or simply reduced sputtering (runs {\tt f-nosp)} and {\tt f-sp.2}), in which the ratio increases less. The latter run could be also more consistent than the fiducial one with recent estimates of dust content in clusters at low redshift (see Section \ref{sec:obs}). On average, the total dust surviving in {\tt f-sp.2} is a factor of 3 greater than in {\tt fid}. This dust excess is almost entirely due to the increased survival of large grains in the hot ICM, while small grains do not deviate significantly in the two cases. This could seem at first sight unexpected, since sputtering affects more promptly small grains than large ones (Eq. \ref{eq:tausp}). What happens is that increasing its timescale by a factor of 5 impacts on the survival of large grains, but the process remains still sufficiently effective to obliterate almost entirely small ones in the hot ICM. Consequently, in the hot environment we predict that the small-to-large grain ratio is affected considerably. In both {\tt fid} and {\tt f-sp.2}, the small grain content shown in \fig\ref{fig:allruns} comes from regions characterized by gas temperature $\lesssim 10^6$ K.



\begin{figure*}
\centering
\includegraphics[width = \columnwidth]{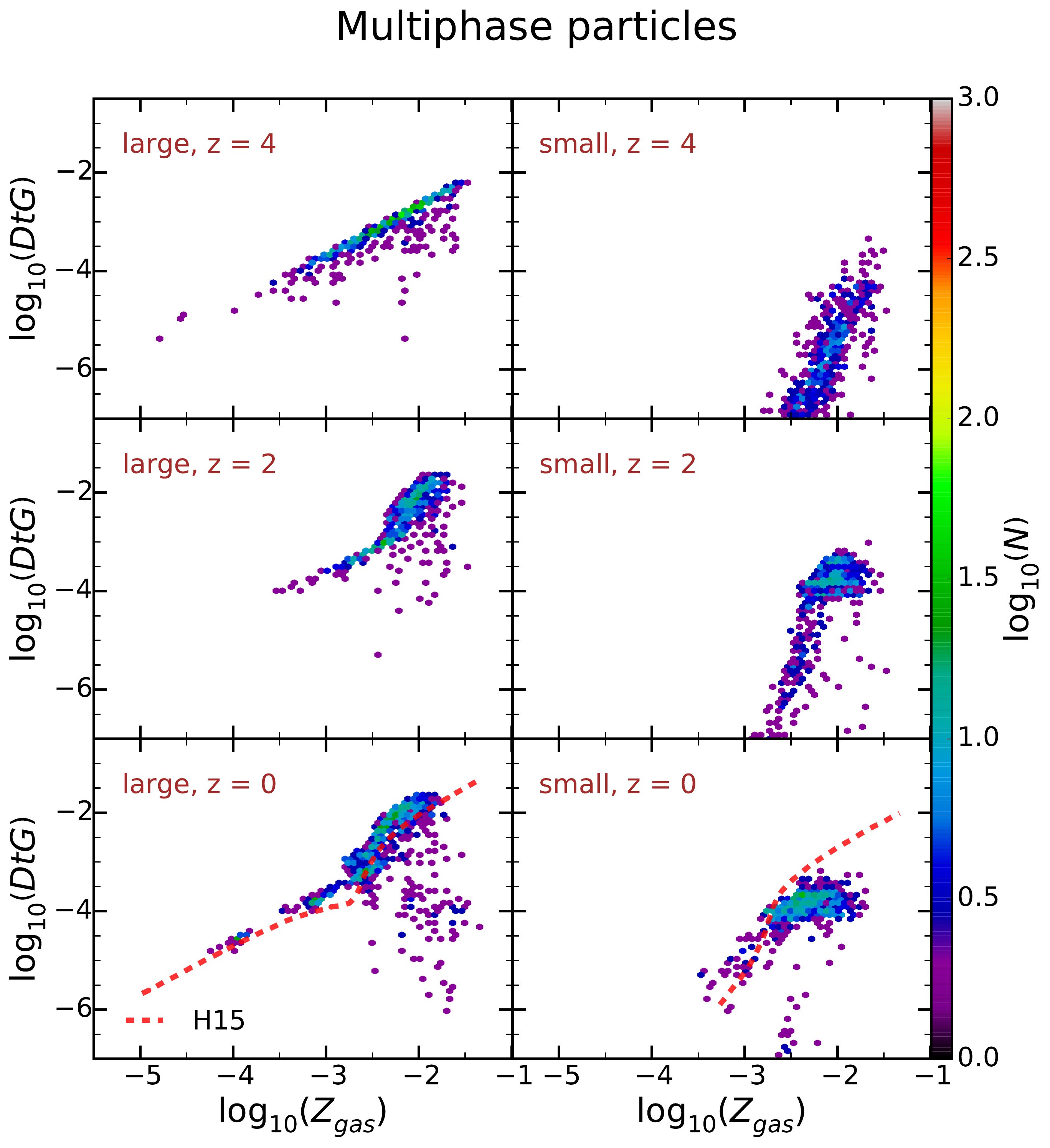}
\includegraphics[width = \columnwidth]{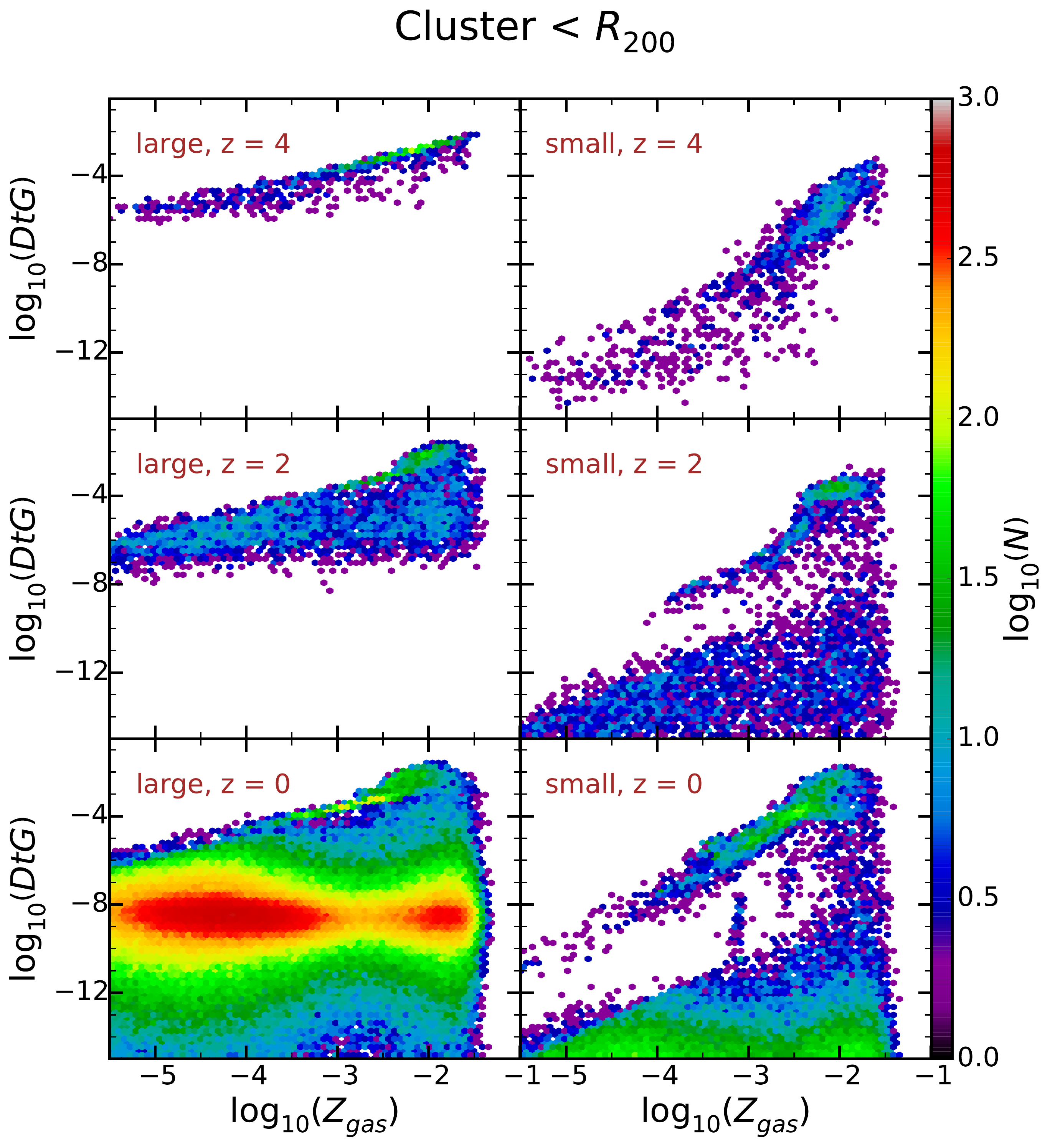}

\caption{2D histogram of the DtG vs gas metallicity for multiphase star forming gas particles  (left) and for all gas particles (right) in the standard run {\tt fid} within $R_{200}$. The black lines represent the relationship predicted by the one-zone model by Hirashita (2015). }
\label{fig:histhex}
\end{figure*}


\subsubsection{DtG and metallicity}

\fig\ref{fig:histhex} presents 2D histograms of the DtG vs gas metallicity $Z_{gas}$ at $z=4$, 2, and 0 for gas particles which are either multiphase in the entire simulated region (left) or  within $R_{200}$ of the main cluster (right). In both cases we used the {\tt fid} run. 

Looking at the multiphase star forming particles, we notice that at an early time of $z=4$, the gas particles cluster along a linear relationship between DtG of large grains and $Z$. Large grains dominate the dust content so this also represents the total behaviour. In this initial phase, the often adopted approximation of a constant ratio between dust and gas metal content is qualitatively good enough over the entire metallicity range of star forming particles. Note that the normalization of this linear relationship between DtG and Z is lower by a factor $\sim 3$ than that generally adopted (i.e.\ DtG $\simeq 0.01$ for $Z \simeq Z_\odot$). The linearity arises because initially the dominating mechanism responsible for the presence of dust in the system is production by stars, for which the basic assumption is that a certain fraction of produced metals goes to the solid state rather than gaseous form.  As time goes on, the ISM evolution processes boost the dust content for a given metallicity. Accretion of gas metals onto small grains provides sufficient mass to enhance the coagulation of small grains
onto large ones. Hence, coagulation dominates over shattering in dense regions. This occurs only above a critical $\log Z \simeq -2,5$ for the fiducial parameters, producing a more than linear increase of DtG with $Z$. However, at still 2-3 times higher $Z$, the DtG slope slows down again toward a linear relationship, now featuring a normalization close to that given by standard MW values. 



As noticed by \citet[][see their figure 7]{aoyama17}, the overall result of this evolution is that at late time $z\le 2$ multiphase particles tend to concentrate in a region whose shape resembles the line along which  the one-zone model by \cite{hirashita15} {\it evolves in time}, shown in the figure with a black line. However when evaluating this result one should keep in mind the very different nature of one-zone computations and simulations. In the latter the gas and star density fields  are sampled with a certain resolution. In the one-zone model, both quantities reported in the 2D histograms of \fig\ref{fig:histhex} are a function of time. In other words, the system moves with time along the line. On the contrary, in the simulation at any given time the gas particles have a broad distribution in the plane, albeit with well defined concentrations. 

The right panels of \fig\ref{fig:histhex} shows that when the 2D distributions are computed for the main cluster of the region and including also non multiphase particles, a population of particles characterized by very low values of DtG shows up. This population becomes more and more dominant over time, because it is produced by sputtering which efficiently destroy dust grains in the hot ICM at $T\gtrsim 10^6$ K.


\subsection{Observational consistency}
\label{sec:obs}

\begin{figure}
\centering
\includegraphics[width = \columnwidth]{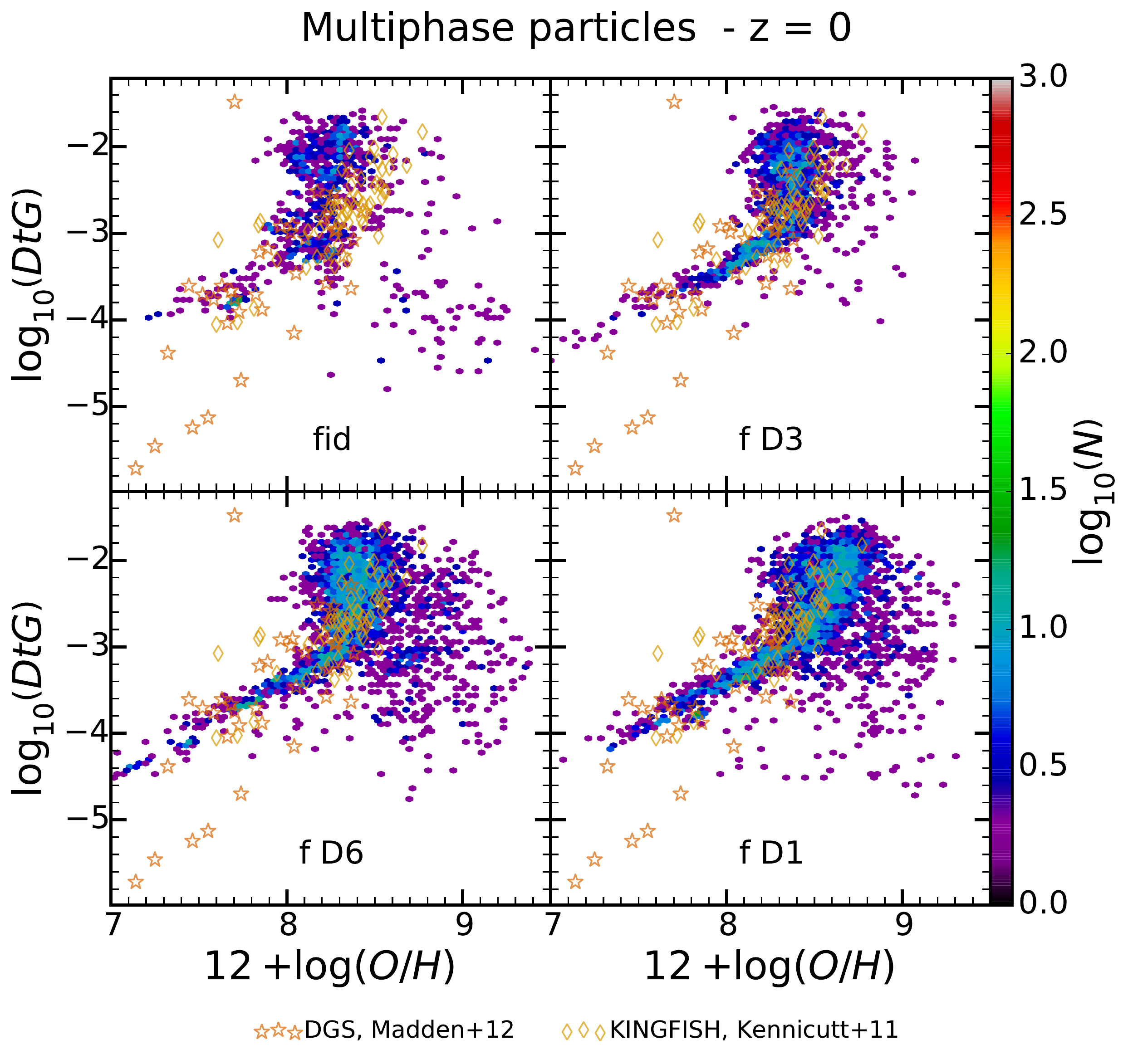}
\caption{2D histogram of the DtG vs Oxygen abundance at $z=0$ compared with galactic data from \protect\citet{kennicutt11} (red diamonds) and \protect\cite{madden12} (black stars).}
\label{fig:histhexs}
\end{figure}

\subsubsection{Dust abundance vs metallicity}
In  \fig\ref{fig:histhexs} we show the 2D distribution of DtG vs (gas) Oxygen abundances of star forming multiphase particles at $z=0$, for the four regions on which we run our fiducial model. This information is similar to that already reported in \fig\ref{fig:histhex}, but now for all the regions considered in the present work, and in a form more directly comparable with observations. We include in the figure data on nearby galaxy samples from \cite{kennicutt11} and \cite{madden12}. It is apparent the broad similarity between the distribution of simulated and the data points.
Note however that the former refers to individual particles residing in star forming regions rather than entire unresolved galaxies, which could explain the larger dispersion. In any case, it is reassuring that the model reproduces reasonably well the observed trend of dust abundance with metallicity.

\begin{figure}
\centering
\includegraphics[width = \columnwidth]{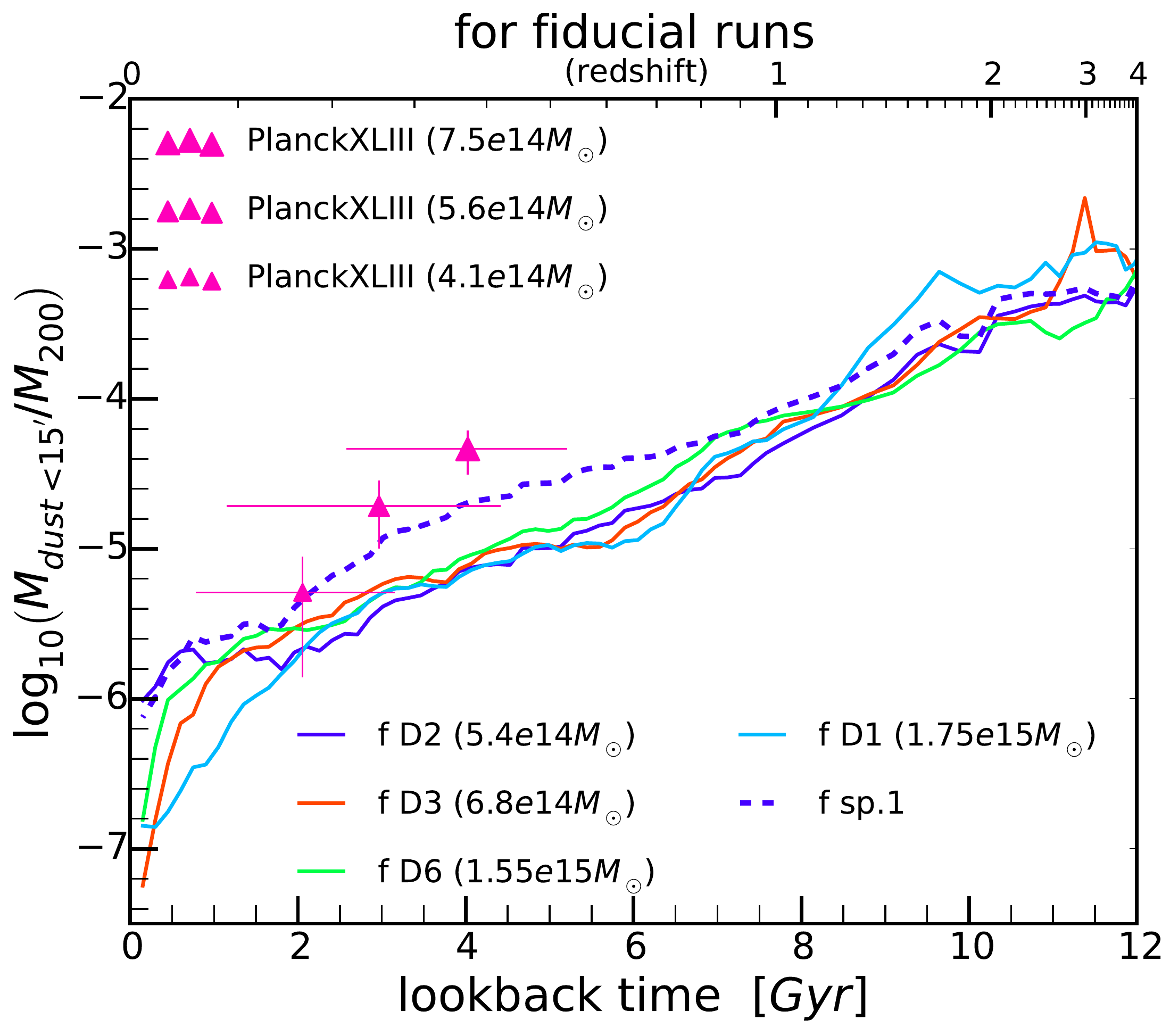}
\caption{For the the 4 different zoom-in cluster simulations  considered in this work ({\tt fid (D2), f-D3, f-D1, f-D6}) we show the evolution of the ratio between the dust mass within a 15 arcmin aperture and $M_{200}$. For the region D2, we show also the run with reduced sputtering, {\tt f-sp.2}. These are compared with the same quantity as estimated in Planck Collaboration (XLIII) et al. (2016), for the whole sample (central point) and for the subsamples at $z<0.25$ and $z>0.25$. The error-bars represent dispersions.}
\label{fig:fidrun}
\end{figure}

\subsubsection{Dust content at low redshift}
In this section we briefly compare with the still limited reliable detections of global dust content in galaxy clusters. \cite{gutierrez17} analyzed 327 clusters of galaxies in the redshift range 0.06-0.70, using maps and catalogs from the Herschel MerMES project. They reported average integrated fluxes of 118.2, 82.3 and 38.0 mJy within 5 arcmin of the cluster centers at 250 $\mu$m, 350 $\mu$m and 500 $\mu$m respectively. Adopting their same assumptions on the dust optical  properties and temperature (essentially the same as the "average" MW dust), these fluxes translate to a total dust mass of $1.7 \times 10^9$ M$_\odot$. Given that their average cluster mass is $1.1 \times 10^{14}$ M$_\odot$ within the same radius, the fraction of dust mass turns out to be about $1.5 \times 10^{-5}$. 
For the main clusters in our four regions we predict an average fraction in the same redshift range of $0.25 \times 10^{-5}$. Thus we are under-predicting the dust content by a factor $\sim 6$. On one hand this discrepancy could be due at least in part to the quite strong and uncertain assumptions entering into the masses estimated from observations, such as the dust optical properties and (constant) temperature. 
In particular, they adopted a long wavelength emissivity that scales as $\lambda^{-\beta}$, with $\beta=2$, which is the standard result of first principle computations of dust opacities by \cite{draine84}. There are several indications that dust emission is better represented at long $\lambda$ by a shallower index $\beta \sim 1.5$, when it is described by single $T$ blackbody \citep[e.g.][]{22planck15}. From a physical point of view, this could be actually the result of a spread in dust temperature. In any case, adopting a smaller $\beta$ in interpreting observed fluxes would lead to smaller dust masses in better agreement with our result. On the other hand the dust content at low redshift in our model clusters is strongly dictated by the sputtering efficiency. For instance, assuming a timescale longer \rev{than that favored by existing literature \citep[e.g.][and references therein]{tsai95}} by a factor 5 (Eq. \ref{eq:tausp}), the final dust content increases by a factor $\sim 3$ (See Fig. \ref{fig:allruns}).

\citet{43planck16} performed a stacking analysis of several hundreds of clusters, wherein IRAS and Planck data are combined to provide a well sampled FIR-submm average SED of clusters. Then they fit this SED with modified black body models to derive dust masses and temperature simultaneously. For the whole sample, whose mean redshift is 0.26 and mean total mass is $M_{200} =5.6 \times 10^{14}$ M$_\odot$, the estimated dust mass is $1.1 \times 10^{10} $ M$_\odot$, adopting their preferred emissivity index $\beta=1.5$. Strictly speaking their dust mass refers to the radius of 15 arcmin within which the fluxes have been integrated. 


In Figure \ref{fig:fidrun}, the time evolution of the ratio $M_{dust}(r < 15')/ M_{200}$ is shown individually for the 4 clusters simulated with the fiducial set of parameters. This is compared with the same ratio as derived from these Planck data.
The comparison is shown not only with the full sample, but also for the two subsample at $z\le 0.25$ and $z>0.25$, comprising 307 and 254 clusters respectively. This figure confirms that our fiducial models under-predict the dust content of the clusters, albeit by a smaller factor $\sim 3$. As such, the model with increased sputtering timescale turns out to be in reasonable agreement with the data. 

\/*
\begin{itemize}
\end{itemize}




At 15 arcmin the radius at redshift  0.4  is  2338.286  kpc.
r  2356.992
M200 at redshift  0.4  of run  run22_D1_stn  is  31712.478  x 10e14 solar masses.
The gas mass enclosed within R200 at redshift  0.4  of run  run22_D1_stn  is  1.35193223741  x 10e14 solar masses.
The dust mass enclosed within  2356.99  kpc at redshift  0.4  of run  run22_D1_stn  is  1.36560880476  x 1e10 solar masses.
The Mdust-to-M200 enclosed within  2356.99  kpc at redshift  0.4  of run  run22_D1_stn  is  4.30621916318e-05
At 15 arcmin the radius at redshift  0.3  is  1805.2161  kpc.
r  1689.682
M200 at redshift  0.3  of run  run22_D6_stn  is  34740.617  x 10e14 solar masses.
The gas mass enclosed within R200 at redshift  0.3  of run  run22_D6_stn  is  1.38022976345  x 10e14 solar masses.
The dust mass enclosed within  1689.68  kpc at redshift  0.3  of run  run22_D6_stn  is  0.618755982982  x 1e10 solar masses.
The Mdust-to-M200 enclosed within  1689.68  kpc at redshift  0.3  of run  run22_D6_stn  is  1.7810736723e-05
At 15 arcmin the radius at redshift  0.3  is  1805.2161  kpc.
r  1689.682
M200 at redshift  0.3  of run  run22_D3_stn  is  34740.617  x 10e14 solar masses.
The gas mass enclosed within R200 at redshift  0.3  of run  run22_D3_stn  is  0.805994940864  x 10e14 solar masses.
The dust mass enclosed within  1689.68  kpc at redshift  0.3  of run  run22_D3_stn  is  0.42002821962  x 1e10 solar masses.
The Mdust-to-M200 enclosed within  1689.68  kpc at redshift  0.3  of run  run22_D3_stn  is  1.20904075947e-05
At 15 arcmin the radius at redshift  0.3  is  1805.2161  kpc.
r  1689.682
M200 at redshift  0.3  of run  run22_stn  is  34740.617  x 10e14 solar masses.
The gas mass enclosed within R200 at redshift  0.3  of run  run22_stn  is  0.543906656901  x 10e14 solar masses.
The dust mass enclosed within  1689.68  kpc at redshift  0.3  of run  run22_stn  is  0.232723707126  x 1e10 solar masses.
The Mdust-to-M200 enclosed within  1689.68  kpc at redshift  0.3  of run  run22_stn  is  6.6988938949e-06
output: (0) DtG    (1) Mdust / M200    (2) M200    (3) Mdust    (4) Mgas
At 15 arcmin the radius at redshift  0.3  is  1963.296  kpc.
r  1837.645
M200 at redshift  0.3  of run  run22_stn  is  32675.966  x 10e14 solar masses.
The gas mass enclosed within R200 at redshift  0.3  of run  run22_stn  is  0.497965596517  x 10e14 solar masses.
The dust mass enclosed within  1837.65  kpc at redshift  0.3  of run  run22_stn  is  0.240443295075  x 1e10 solar masses.
The Mdust-to-M200 enclosed within  1837.65  kpc at redshift  0.3  of run  run22_stn  is  7.35841428758e-06
At 15 arcmin the radius at redshift  0.2  is  1306.4819  kpc.
r  1128.8
M200 at redshift  0.2  of run  run22_stn  is  43024.786  x 10e14 solar masses.
The gas mass enclosed within R200 at redshift  0.2  of run  run22_stn  is  0.672263726128  x 10e14 solar masses.
The dust mass enclosed within  1128.8  kpc at redshift  0.2  of run  run22_stn  is  0.135747881399  x 1e10 solar masses.
The Mdust-to-M200 enclosed within  1128.8  kpc at redshift  0.2  of run  run22_stn  is  3.15510881098e-06

*/

\section{Summary and future prospects} \label{sec:conclusions}
In this work we have introduced a state of the art treatment of dust production and evolution
in our version of the simulation code {\footnotesize GADGET-3}. We take advantage of the detailed description of chemical evolution already included in our version of {\footnotesize GADGET-3} to trace separately the two dust species that are believed to populate the ISM, namely carbonaceous and silicate grains. We also trace at a basic level the continuum size distribution of grains by means of the two-grain-size approximation introduced and tested by \cite{hirashita15}. 
In our code, large (nominally $0.1 \mu$m) dust grains are originated by simulation stellar particles from three stellar channels, AGB winds, core collapse SNae, and SNIa. These grains are spread along with metals to the surrounding SPH gas particles, where we allow for various ISM evolution processes affecting dust properties to occur. 
In brief, large grains are shattered onto small (nominally $0.01 \mu$m) grains if the gas density is low enough. On the other hand, in dense star forming SPH gas particles metals accrete onto small grains and small grains coagulate onto large grains. We also take into account dust destruction by SN shocks and by sputtering in the hot ($T \gtrsim 10^6$ K) ICM. We evaluate timescales for the above mentioned ISM processes as a function of the physical conditions of the SPH particles, and within each we evolve the dust and gas metal contents.  As remarked above, the former is divided into 4 types, namely large carbon, small carbon, large silicates, and small silicates.



As a first test, we apply the method to cosmological zoom-in simulations of four massive ($M_{200} \geq 3 \times 10^{14} M_{\odot}$) galaxy clusters. 
During the early stages of assembly of the cluster at  $z \gtrsim 3$,  where the star formation activity is at its maximum in our simulations, the proto-cluster regions are rich of dusty gas (\fig\ref{fig:distrclus}). Compared to runs in which only dust production in stellar ejecta is active, runs including processes occurring in the cold ISM enhance the dust content by a factor $2-3$.
However, the dust properties in this stage turn out to be significantly different than those observationally derived for the {\it average} Milky Way dust, and commonly adopted in calculations of dust reprocessing (\fig\ref{fig:distrsilc}). 
We stress that this results is not unexpected.
We show that these differences may have a strong impact on the predicted spectral energy distributions (\fig\ref{fig:skirtseds}). At low redshift our model reproduces reasonably well the trend of dust abundances over metallicity as observed in local galaxies (\fig\ref{fig:histhexs}). However we under-produce by a factor of 2 to 3 the total dust content of clusters estimated observationally at low-z $ \lesssim 0.5$ using IRAS, Planck and Herschel satellites data. This discrepancy can be solved by decreasing the efficiency of sputtering which erodes dust grains in the hot ICM (\fig\ref{fig:fidrun}).



The most immediate purpose of our effort is to have enough information on the simulated ISM/ICM to compute observational properties of simulated objects by means of radiative transfer post processing. This is in line with what we already achieved in \cite{granato15}, but with less assumptions required on dust properties. However this work can be also regarded as a first step in the direction of a more sophisticated prescriptions for the sub-resolution physics. For instance, it will be possible to estimate the contribution of dust to the formation of molecular gas. This estimate is a primary ingredient in more advanced implementations of star formation in simulations \citep[e.g. MUPPI,][]{murante15}.


\section*{Acknowledgements}
\rev{We are indebted to the anonymous referee for the very careful reading of our manuscript and several important suggestions that improved its quality.} 
We thank Volker Springel for providing the non-public version of the GADGET3 code, and Peter Camps for prompt assistance and advice in using the radiative transfer code SKIRT.
EG is supported by the INAF 2015 PhD grant "Galaxy clusters in their infancy: confronting simulated and observed IR/sub-mm properties". 
This project has received funding from the European Union's Horizon 2020 Research and Innovation Programme under the Marie Sklodowska-Curie grant agreement No 73437
The simulations were carried out at the following facilities (i) Mendieta Cluster from CCAD-UNC, which is part of SNCAD-MinCyT, Argentina; (ii) European Exascale System Interconnect and Storage (iii)  CINECA (Italy), with CPU time assigned through ISCRA proposals and an agreement with the University of Trieste. The post-processing and analysis has been performed using the PICO HPC cluster at CINECA through our expression of interest. This publication has received funding from the European Union's Horizon 2020 research and innovation program under grant agreement No 730562 [RadioNet]

The analysis was conducted using IPython2 \citep{perez07}, SciPy3 \citep{jones01}, NumPy4 \citep{vanderwalt11}, and MatPlotLib5 \citep{hunter07}.




\bibliographystyle{mnras}
\bibliography{bibliosmall}

\label{lastpage}
\end{document}